\documentclass[useAMS,usenatbib,referee]{biom}
\usepackage{url} 
\usepackage{hyperref} % Added for proper referencing
\usepackage{amsmath}
\usepackage{mathrsfs}
\usepackage{xcolor}
\usepackage[figuresright]{rotating}

\def\bSig\mathbf{\Sigma}

\title{Survival Analysis with Discrete Biomarkers Under a Semiparametric Bayesian Conditional Poisson Model}

\author{Aijun Yang$^{1}$, Phineas T. Hamilton$^{2}$, Brad H. Nelson$^{2}$, Julian J. Lum$^{2,3}$\\ \textbf{Mary Lesperance}$^{\mathbf{1}}$\textbf{, and} \textbf{Farouk S. Nathoo}$^{\mathbf{1},\mathbf{*}}$\\
$^{1}$Department of Mathematics \& Statistics, University of Victoria, BC, V8P5C2, Canada\\
$^{2}$Deeley Research Center, BC Cancer Agency, Victoria, BC, V8R6V5, Canada \\
$^{3}$Department of Biochemistry and Microbiology, University of Victoria, BC, V8P5C2, Canada}

\email{nathoo@uvic.ca}

\begin{document}

%  This will produce the submission and review information that appears
%  right after the reference section.  Of course, it will be unknown when
%  you submit your paper, so you can either leave this out or put in 
%  sample dates (these will have no effect on the fate of your paper in the
%  review process!)

\date{{\it Received xxx} 2025. {\it Revised xxx} 2025.  {\it
Accepted xxx} 2025.}

%  These options will count the number of pages and provide volume
%  and date information in the upper left hand corner of the top of the 
%  first page as in published papers.  The \pagerange command will only
%  work if you place the command \label{firstpage} near the beginning
%  of the document and \label{lastpage} at the end of the document, as we
%  have done in this template.

%  Again, putting a volume number and date is for your own amusement and
%  has no bearing on what actually happens to your paper!  

\pagerange{\pageref{firstpage}--\pageref{lastpage}} 
\volume{xx}
\pubyear{2025}
\artmonth{xxx}

%  The \doi command is where the DOI for your paper would be placed should it
%  be published.  Again, if you make one up and stick it here, it means 
%  nothing!

\doi{10.1111/j.1541-0420.2005.00454.x}

%  This label and the label ``lastpage'' are used by the \pagerange
%  command above to give the page range for the article.  You may have 
%  to process the document twice to get this to match up with what you 
%  expect.  When using the referee option, this will not count the pages
%  with tables and figures.  

\label{firstpage}

%  put the summary for your paper here

\begin{abstract}
Discrete biomarkers derived as cell densities or counts from tissue microarrays and immunostaining are widely used to study immune signatures in relation to survival outcomes in cancer. Although routinely collected, these signatures are not measured with exact precision because the sampling mechanism involves examination of small tissue cores from a larger section of interest. We model these error-prone biomarkers as Poisson processes with latent rates, inducing heteroscedasticity in their conditional variance. While critical for tumor histology, such measurement error frameworks remain understudied for conditionally Poisson-distributed covariates. To address this, we propose a Bayesian joint model that incorporates a Dirichlet process (DP) mixture to flexibly characterize the latent covariate distribution. The proposed approach is evaluated using simulation studies which demonstrate a superior bias reduction and robustness to the underlying model in realistic settings when compared to existing methods. We further incorporate Bayes factors for hypothesis testing in the Bayesian semiparametric joint model. The methodology is applied to a survival study of high-grade serous carcinoma where comparisons are made between the proposed and existing approaches. Accompanying R software is available at the GitHub repository listed in the Web Appendices.
\end{abstract}

%  Please place your key words in alphabetical order, separated
%  by semicolons, with the first letter of the first word capitalized,
%  and a period at the end of the list.
%


%  As usual, the \maketitle command creates the title and author/affiliations
%  display 

\maketitle

%  If you are using the referee option, a new page, numbered page 1, will
%  start after the summary and keywords.  The page numbers thus count the
%  number of pages of your manuscript in the preferred submission style.
%  Remember, ``Normally, regular papers exceeding 25 pages and Reader Reaction 
%  papers exceeding 12 pages in (the preferred style) will be returned to 
%  the authors without review. The page limit includes acknowledgements, 
%  references, and appendices, but not tables and figures. The page count does 
%  not include the title page and abstract. A maximum of six (6) tables or 
%  figures combined is often required.''

%  You may now place the substance of your manuscript here.  Please use
%  the \section, \subsection, etc commands as described in the user guide.
%  Please use \label and \ref commands to cross-reference sections, equations,
%  tables, figures, etc.
%
%  Please DO NOT attempt to reformat the style of equation numbering!
%  For that matter, please do not attempt to redefine anything!

\section{Introduction}
\label{s:intro}
Covariate measurement error is an important issue in regression analysis for survival data that can lead to bias in estimated hazard ratios, inflation of variance, and produce misleading clinical or epidemiological conclusions \citep{Carroll2006,Prentice1982, Gustafson2003,Fuller1987,Yi2021}. While methodologies for addressing errors in continuous covariates are well-developed, discrete surrogates in the form of count data, particularly those modeled as conditionally Poisson-distributed variables, have received far less attention. This issue is especially important in cancer research, where discrete biomarkers derived from cell counts (e.g., immune cell densities in tumor tissue microarrays [TMAs]) act as proxies for unobservable densities across larger tissue areas.

Tissue microarrays support high-throughput histological profiling in cancer research by enabling efficient quantification of biomarkers from limited tissue specimens. However, the use of small, subsampled tissue cores introduces heteroscedastic non-Gaussian non-additive measurement error in count-based biomarkers such as tumor-infiltrating lymphocytes (TILs). These complexities pose challenges for survival analysis aiming to evaluate associations between immune cell infiltration and clinical outcomes, where accurate covariate measurement is critical. The issue is relevant in tumor histology studies of high-grade serous carcinoma (HGSC), a highly aggressive gynecologic malignancy characterized by early dissemination and rapid emergence of treatment resistance \citep{Havasi2023}. Measurement error stemming from technical variability and TMA-specific subsampling can attenuate effect estimates, diminish statistical power, and compromise inferential validity \citep{Yang2024}. These limitations highlight the need for statistical methods that explicitly account for the discrete, non-Gaussian, and non-additive error structures in tumor histological biomarkers.

Frequentist methodologies for measurement error correction have been developed extensively. Regression calibration (RC), introduced by \citet{Prentice1982}, replaces mismeasured covariates with their conditional expectations. \citet{Wang2008} refined RC for the Cox model, demonstrating robustness under small errors. However, RC deteriorates with large or nonlinear errors and requires internal validation data \citep{Spiegelman1997}. For the studies we are concerned with, internal validation data is not available.

The corrected score function approach, pioneered by \citet{Nakamura1992} for the Cox proportional hazards (PH) model, modifies estimating equations to account for bias induced by mismeasured covariates. This method assumes additive error structures and requires knowledge of the error variance. It was later extended to time-dependent covariates by \citet{Kong1999}. \citet {Augustin2004} applied Nakamura’s methodology to the Breslow likelihood, which is equivalent to the partial likelihood in the absence of measurement error. \citet{Li2006} proposed a method to estimate regression coefficients by solving an estimating equation, averaging partial likelihood scores from imputed true covariates. \citet{Yi2007} advanced these techniques with conditional estimating equations and \citet {Yan2015} discussed a corrected profile likelihood method for the Cox model. More recently, \citet{Cao2022} proposed an approximate profile likelihood estimation approach, offering an alternative to the corrected score method. Although regression calibration and corrected score methods offer corrections for measurement error, their utility is often limited by reliance on restrictive assumptions such as additive error structures and constant error variance. These methods are further constrained by sensitivity to large or nonlinear errors and they typically require external validation data, reducing their robustness and feasibility in settings characterized by complex error mechanisms or limited auxiliary information.

Simulation-extrapolation (SIMEX), proposed by \citet{Cook1994}, simulates measurement error-inflated datasets to extrapolate bias-corrected estimates. Extensions include the incorporation of heteroscedastic errors \citep{Devanarayan2002} and errors with non-zero mean \citep {Parveen2021}. SIMEX has been applied to accelerated failure time (AFT) models \citep{Greene2004,He2007,He2012,Zhang2014} and later to the Cox PH  model \citep{simex2019}. Recent advances address high-dimensional data via graphical PH models \citep{Chen2020} and SIMEXBoost with Lasso regularization to incorporate variable selection \citep{Chen2024}. As one of our comparative approaches, we extend the SIMEX algorithm to the conditional Poisson context (POI-Gamma-SIMEX).

Bayesian methods are typically based on joint modeling formulations, allowing for the incorporation of prior knowledge, quantification of uncertainty, and finite sample inference. \citet{Richardson1993} pioneered data augmentation for measurement error based on MCMC sampling. \citet{Ibrahim2001} extended this to survival models, integrating the Weibull and Cox PH models with error correction. For nonparametric error modeling, Dirichlet Process Mixtures (DPMs) flexibly model latent covariate distributions, outperforming parametric methods under non-Gaussian errors \citep{Gelman2013, Sarkar2014}. This approach leverages the flexibility of DPMs to model latent covariate distributions adaptively, circumventing parametric assumptions \citep{Ferg1973,Antoniak1974}. While DP mixtures have been applied to continuous measurement error problems \citep{Muller2015, Pan2022}, their potential for discrete covariates remains underexplored. \citet{Sinha2017} linked longitudinal covariates to Cox PH models, addressing time-dependent errors, while \citet{Tang2022} incorporated heteroscedastic errors into Bayesian AFT models.

Motivated by the measurement error in biomarkers arising from tumor histology studies of HGSC, we propose a Bayesian semiparametric joint model for conditionally Poisson-distributed biomarkers in survival analysis. The Bayesian joint model integrates a nonparametric Dirichlet Process (DP) mixture of gamma distributions within a conditional Poisson model and a Cox proportional hazards (PH) framework to jointly model survival and discrete biomarkers. In addition, the model can be applied with biomarkers that are observed without replication. Our work is distinct from \citet{Pan2022} who incorporated a Dirichlet process mixture of Gaussians to model additive non-Gaussian measurement errors in Cox regression with repeated surrogates, and who also considered binary covariates subject to misclassifications. To our knowledge, there is limited methodology dealing with conditionally Poisson surrogates, with the exception of methods for linear regression \citep{Li2004poi,Yang2024}. 

In addition to the semiparametric joint model, we propose the use of Bayes factors for hypothesis testing \citep{KassRaftery1995}. We compute Bayes factors from the posterior samples using the Savage-Dickey density ratio \citep{Wagenmakers2010} which is applicable under our assumed prior distribution. We evaluate the performance of Bayes factors in the setting of our model using both simulations as well as in our application to HGSC. In the latter case, the inference from Bayes factors and p-values is contrasted, and an inferential paradox is noted \citep{Lindley1957}. Subsequently, the posterior distribution, effect sizes and the sensitivity to prior distributions are carefully examined.

The remaining sections of the paper proceed as follows. Section 2 presents the model development and implementation strategies. We evaluate the performance of our method through simulations in Section 3 and demonstrate their application in Section 4. Finally, Section 5 concludes with a discussion of our results and avenues for future work.

\section{Methodology}
\label{s:method}
We introduce our DPM joint model and describe its implementation details. In addition, as one comparative approach, we extend the standard SIMEX algorithm to the conditional Poisson setting. %\textcolor{blue}{Aijun, please go through the document and make sure all vectors are written in bold font.}
\subsection{The Cox Model with a Conditional Poisson Error Model for Biomarkers}
For each subject, $i=1,...,n$, let $T_i, C_i, \delta_i$ and $\mathbf{Z_i^*}$ represent the survival time, censoring time, event indicator $\delta_i=
I(T_i \le  C_i )$, and a covariate vector of dimension $J$, respectively. We observe $U_i$=min$(T_i,C_i)$. Let $Y_i(t)=I(U_i\ge t)$ be the indicator process that subject $i$ is at risk at time $t$. Under the Cox model \citep{Cox1972}, the hazard function is expressed as 
\begin{equation}
\label {eq1}
h_i(t;\mathbf{Z_i^*})=h_0(t) \exp \{\boldsymbol{\beta}^{'}\mathbf{Z_i^*}\}
\end{equation}
where $h_0(t)$ is the baseline baseline hazard function and $\boldsymbol{\beta}$ is a J-vector containing unknown regression parameters. We use a flexible piecewise constant form for the baseline hazard function where $h_0(t) = h_{l}$ for $t \in [\tau_l,\tau_{l+1})$, $l=0,...,m$, where the intervals are based on a set of $m$ inner knots $\tau_{0}=0<\tau_{1}< \dots < \tau_{m} < \tau_{m+1} = \infty$. 

%he partial likelihood function for $\boldsymbol{\beta}$ is
%\begin{equation*}
%\mathrm{L}(\boldsymbol{\beta\lvert(U,Z^*,\delta)}) = \prod_{i=1}^{n}\{\frac{\exp\{\beta^{'}Z_i^*%(U_i)\}}{\sum_{j}^{n}Y_j(U_i) \exp\{\beta^{'}Z_j^*(U_i)\}}\}^{\delta_i}
%\end{equation*}
%The partial log-likelihood is
%\begin{equation}
%\label {eq2}
%\begin{eqnarray*}
%\boldsymbol{\ell(\beta\lvert(U,Z,\delta))}=\sum_{i=1}^{n}{\delta_i\left[\beta^{'}Z_i^*(U_i)-%\log(\sum_{j}^{n} Y_j(U_i)\exp(\beta^{'}Z_j^*(U_i)))\right] }
%\end{eqnarray*}
%\end{equation}
%For simplicity, we assume the covariates are time independent. So $Z_i^*(U_i)$ will be just $Z_i^*$. 
We let $\mathbf{Z^*=(X,Z)}^T$, where $\mathbf{Z}$ are assumed to be measured without error, and where $X$ is the biomarker of interest, typically a surrogate for true cell density $W_T/A_T$ over a larger tissue area $A_T$, which cannot be measured directly from a small core sampled from that area. Instead, the variable $W$, representing the biomarker cell count of the selected TMA core with area $A$ is measured. The estimated cell density is $W/A$ and is used as a surrogate for $X$  in the regression model. 
The random cell locations within a tissue sample are assumed to be distributed according to a homogeneous Poisson process with intensity $X = \frac{W_T}{A_T}$. 
Under the Poisson process model for the biomarker cell distribution, we have for subject $i$,
%\begin{equation}
%\label{eq0}
\[
W_i\lvert X_i \overset{\displaystyle \text{ind}}{\sim} \text{Poisson} (X_i A_i) \ \ i=1,...,n, 
%\end{equation}
\]
where $X_i$ is the unobserved true tissue cell density for the $i^{th}$ subject and $W_i$ is the observed number of cells on the core with area $A_i$. The true density is incorporated as a covariate in (\ref{eq1}).

%Thus, $E[W_i\lvert X_i]= Var [W_i\lvert X_i]=A_i X_i$, and $E[W_i]=A_i E[X_i]$. We note that the model is Poisson conditional on $X_{i}$, and therefore overdispersion in the observed $W_{i}$ does not preclude the use of this model. 
\subsection{A Semiparametric Bayesian Model for Biomarker Densities} \label {subDP}
The true cell densities $X$ are modeled with a Dirichlet Process (DP) mixture of gamma distributions
\[
X_i \lvert a_i, b_i \stackrel{ind}{\sim} \text{Gamma} (a_i,b_i),\,\, a_i,b_i | P_{a,b} \stackrel{iid}{\sim} G_{a,b},\,\, G_{a,b} \sim DP(\alpha,G_{0}) \text{,}
\]
where the base measure $G_0$ is taken as $\text{Gamma}(a_0,\eta_0) \times \text{Gamma}(b_0,\gamma_0)$
where $a_0,\eta_0,b_0,\gamma_0$ are fixed and known. 

We employ the stick-breaking representation of the Dirichlet process (see, e.g. \citet{DP1994}, \citet{DP2008}) which expresses the random measure $G \sim DP(\alpha,G_0)$ as
$
G=\sum_{k=1}^\infty \pi_k \delta_{\theta_k}(\cdot)
$
where $\alpha$ is the Dirichlet mass or concentration parameter, $\theta_k$ are independent draws from the base distribution $G_0$, $\delta_{\theta_k}$ is a point mass at $\theta_k$, and $\pi_k$ is the probability mass for the atom $\theta_k$. 
The stick-breaking process for the weights $\pi_k$ is: $\pi_1=\nu_1$, where $\nu_1 \sim Beta(1,\alpha)$ and for $k >1$ we let 
\[
\pi_k=\nu_k\prod_{j=1}^{k-1}(1-\nu_j),\,\, \nu_j \stackrel{iid}{\sim} Beta(1,\alpha).
\]
We interpret $\pi_k$ as the $k$-th segment of a unit length stick, obtained by breaking off a fraction $\nu_k$ of the remaining stick length, $\prod_{j=1}^{k-1}(1-\nu_j)$. The concentration parameter $\alpha$ inversely influences the expected size of each stick: larger $\alpha$ leads to smaller, more numerous components; smaller $\alpha$ leads to larger, fewer components. As an approximation, we truncate the infinite process to have $K<\infty$ components and set $\pi_K=\prod_{j=1}^{K-1}(1-\nu_j)$ so that $\sum_{i=1}^K\pi_i=1$. \citet{Ohlssen2007} proposed a pragmatic method for selecting $K$ by linking it to the concentration parameter $\alpha$ and a small tolerance threshold $\epsilon$, which represents the acceptable expected probability mass allocated to the "tail" (i.e., components beyond $K$). For moderate $\alpha$, the truncation level $K$ can be approximated as $
K \approx 1-\alpha \log{\epsilon}$, ensuring that the cumulative weight of components beyond $K$ is unlikely to exceed $\epsilon$. 
 %Now construct a hierarchical model for estimating 
%true covariate $X_i$, as follows:
%\begin{align*}
%(X_i \lvert L_i, a_{L_i},b_{L_i}) &\sim \text{Gamma}(a_{L_i},b_{L_i}), \\
%L_i \lvert \boldsymbol{\pi} & \sim \text{Categorical } (\pi_1,\pi_2,...), \text{ i.e., } %P(L_i=k)=\pi_k\\
%\pi_i &\sim \text{Stick-Breaking} (\alpha),\\
% (a_{L_i},b_{L_i}) &\sim G_0.
%\end{align*}
%In fact, we can view $L_i$, $\nu_k$, and $(a_{L_i},b_{L_i})$ as latent variables representing cluster %assignment of $X_i$, stick-breaking proportion, and parameters of the Gamma components.
\subsection{The Semiparametric Joint Model for Survival and Biomarker Densities}

The complete model specification for survival and biomarker densities is expressed as
$$
h_i(t) =h_0(t)\exp(\boldsymbol{\beta_z} \mathbf{Z_i} + \beta_x X_i),\,\,\, h_0(t) = h_{l},\,\, t \in [\tau_l,\tau_{l+1}),\,\, l=0,...,m
$$
$$
W_i\lvert X_i \overset{\displaystyle \text{ind}}{\sim} \text{Poisson} (X_i A_i) 
$$
$$
X_i \lvert a_i, b_i \stackrel{ind}{\sim} \text{Gamma} (a_i,b_i),\,\, a_i,b_i | P_{a,b} \stackrel{iid}{\sim} G_{a,b},\,\, G_{a,b} \sim DP(\alpha, G_{0})
$$
$$
G_0 \equiv \text{Gamma}(a_0,\eta_0) \times \text{Gamma}(b_0,\gamma_0)
$$
$$
h_{l} \stackrel{iid}{\sim} \text{Gamma}(a_{h}, b_{h}),\,\, \beta_{z_{j}} \stackrel{iid}{\sim}   N(\mu_{\beta}, \sigma_{\beta}^2),\,\, \beta_{x} \sim N(\mu_{x}, \sigma_{x}^2),\,\, \alpha \sim \text{LN}(\mu_{\alpha},\sigma_{\alpha}^{2})
$$
Specific choices for the prior distributions are based on $a_{h} = b_{h} = 0.01$, $\mu_{x} = \mu_{\beta} = \mu_{\alpha}=0$, $\sigma_{x}^2 = 100$, $\sigma_{\beta}^{2} = \sigma_{\alpha}^{2} =1$, $a_0=b_0=\eta_0=\gamma_0=0.1$,
which correspond to priors that are weakly informative for datasets the size of which we have considered in our simulations and application. These choices ensure that the priors provide sufficient flexibility while avoiding strong influence on the posterior distribution. To clarify, while the posterior distribution is not heavily influenced by the prior distribution for datasets of the size under consideration, Bayes factors can be sensitive to prior distributions even when the posterior distribution is not. In our application, we take advantage of this fact and use a prior sensitivity analysis to glean additional insight into our data.

The likelihood function arises as the product of two terms, the first corresponding to the Cox model for survival, and the second corresponding to the conditional Poisson model for the biomarker. Under a Cox model and a piecewise constant baseline hazard function, the survival component of the likelihood can itself be expressed in the form of a Poisson likelihood with contributions for each interval and an offset term (see e.g., \citet{Bugs2012}, \citet{Ibrahim2001}, \citet{christensen2010}). The posterior distribution is sampled using a Markov chain Monte Carlo algorithm programmed in the rjags package within R \citep{Plummer2019}. This implementation uses the stick-breaking representation of the Dirichlet process with a truncation to $K$ components as discussed in Section \ref{subDP}. Accompanying R software is available at the GitHub repository listed at the end of the article.

\subsection{Bayesian Hypothesis Testing}
Bayesian hypothesis testing using the Bayes factor offers a principled approach to compare the relative evidence for two hypotheses, $H_0$ and $H_1$, by quantifying how strongly the observed data supports one model over the other. The Bayes factor $BF_{10} = P(Y|H_{1})/P(Y|H_{0})$ is the ratio of the marginal likelihoods of the data under $H_1$ and $H_0$, providing a continuous measure of evidence. When hypotheses correspond to nested models, and the prior density of the nuisance parameters under the null hypothesis is the same as the conditional prior density of those parameters under the alternative, conditional on the constraints implied by the null hypothesis, the Savage-Dickey density ratio \citep{Wagenmakers2010} provides a computationally convenient way to compute the Bayes factor as the ratio of the marginal posterior density to the prior density of the parameter under test evaluated at the value specified by $H_0$. This avoids the need to explicitly calculate marginal likelihoods, making it particularly efficient for testing 1-dimensional point-null hypotheses. In our context, interest lies specifically with the test $H_{0}: \beta_{x} = 0$ versus $H_{1}: \beta_{x} \ne 0$ corresponding to the association between the unobserved density of the biomarker and survival. 

We investigate the finite sample properties of the Bayes factor over a range of sample sizes demonstrating its behavior under both the null and alternative (Web Appendix A). The data are generated using a parametric Weibull survival regression model with two covariates. The shape and scale parameters for the Weibull distribution are both set to 1. Under the alternative hypothesis $H_1$, the coefficient for the covariate subject to measurement error is set to $\beta_{x} = 0.5$, while the coefficient for covariate $Z$ is fixed at $\beta_{z} = 0.1$. The sample size $n$ varies across values of $50, 100, 200, 300, 400,$ and $600$. For each sample size, $1,000$ datasets are simulated with no censoring applied. We use a gamma distribution with shape parameter $a = \frac{2}{3}$ and scale parameter $b = 3$ (setting 2 in Section \ref{s:sim}) to simulate covariate $X$. As the sample size increases, the sampling distribution of $BF_{10}$ shows a rapid increase (decrease) under $H_{1}$ ($H_{0}$). 

\subsection{POI-Gamma-SIMEX Estimator for Comparison}
As another approach for comparison, we develop an extension of the SIMEX algorithm \citep{Cook1994} for the semiparametric Cox proportional hazards models and conditionally Poisson covariates. %To use standard SIMEX, you need observed data with error-prone covariates, a known or estimable measurement error variance, a set of simulation levels $(\lambda)$, and a specified model.
SIMEX assumes an additive measurement error model $W = X + U$, where the error $U$ is normally distributed and independent of both the true covariate $X$ and the response $Y$. The method involves simulating and adding additional error at increasing levels of noise $\lambda$, with $\lambda = 0$ corresponding to the observed covariates. Model parameters are estimated across a range of nonnegative $\lambda$ values, and extrapolating back to $\lambda = –1$ to correct for measurement error. Accurate results depend on correct model specification and reliable error variance estimates. From Proposition 2.1 in \citet {Yang2025}, we obtain a strongly consistent estimator for the core-specific measurement error variance $Var[W_{i}/A_{i} - X_{i}]$ as $\hat{\sigma}^2_{e_i}=\frac{\bar{W}}{A_i}$, where $A_i$ is the TMA core area. There, a heteroscedastic standard additive SIMEX approach is applied to the conditionally Poisson case. In contrast to this, we generate mean-zero unit-variance pseudo-errors using an approach that is tailored to the Poisson context where we use pseudo-errors taking the form
\[
\xi_e=Y - u_e, \text{ where } Y \sim \text{Poisson}(u_e) \text{ and } u_e \sim \text{Gamma}(1,1).
\]
This ensures compatibility with the Poisson structure of the measurement error model. Subsequent steps (e.g., parameter estimation via extrapolation) align with the standard SIMEX framework and incorporate the estimated measurement error variance.

%While the POI-Gamma-SIMEX method requires a known or estimable error variance and relies on extrapolation techniques. In contrast, the DP mixture joint model provides a flexible, nonparametric framework that jointly models latent variables and observed data. To better understand the strengths and limitations of each method, we now compare their performance.

\section{Simulation}
\label{s:sim}
Simulation studies are carried out to empirically evaluate the bias and mean squared error (MSE) of competing estimation methods for the Cox proportional hazards regression model with conditional Poisson covariates: (1) the estimator obtained using the true covariate; (2) the naïve estimator obtained using the observed covariate; (3) POI-Gamma-SIMEX; (4) a parametric Bayesian joint model based on a Gamma distribution for the underlying density; and (5) the proposed semiparametric Bayesian joint model estimator. When implementing POI-Gamma-SIMEX we adopt a quadratic extrapolant function. We consider various settings based on the ratio of the true covariate variance to the observed covariate variance, a measure inversely representing the amount of distortion in the observed covariate. We implement the Bayesian approach with $m=5$ equal-length intervals for the baseline hazard and $K=5$ in the Dirichlet process (DP) mixture. %Sensitivity analyses across parameter values are conducted in the application.

The data are simulated using a parametric Weibull survival regression model with two covariates. One unobserved covariate $(X)$ representing biomarker density and the observed count $(W)$ within area $(A)$ follows a Poisson distribution with a mean of $(XA)$. The discrete surrogate for the true covariate is $W/A$. For simplicity, the area is set to $A=1$ in the simulation. The covariate $Z$ is assumed perfectly observed and generated from a standard normal distribution. The shape and scale parameters for the Weibull distribution are both set to 1. The coefficient for the covariate with measurement error is set to $\beta_{x} = 0.5$. The coefficient associated with covariate $Z$ is set to $\beta_{z} = 0.1$. The sample size is set to be $n=100$, and $1,000$ datasets are simulated for each setting both without censoring and with 20\% censoring. We assume $X$ follows a gamma distribution with shape and scale parameters $a$ and $b$ being chosen such that the variance ratio of true covariate and surrogate, $\frac{Var[X]}{Var[W]} = ab^2/(ab+ab^2)$ take values: (1) ratio = 0.9, for a = 0.1, b = 9; (2) ratio=0.75, for a = 2/3, b = 3; (3) ratio = 0.5, for a = 2, b = 1; and (4) ratio=0.5, for a=10, b=1. 
\begin{figure} [!htbp]
\centerline{\includegraphics[scale=0.8]{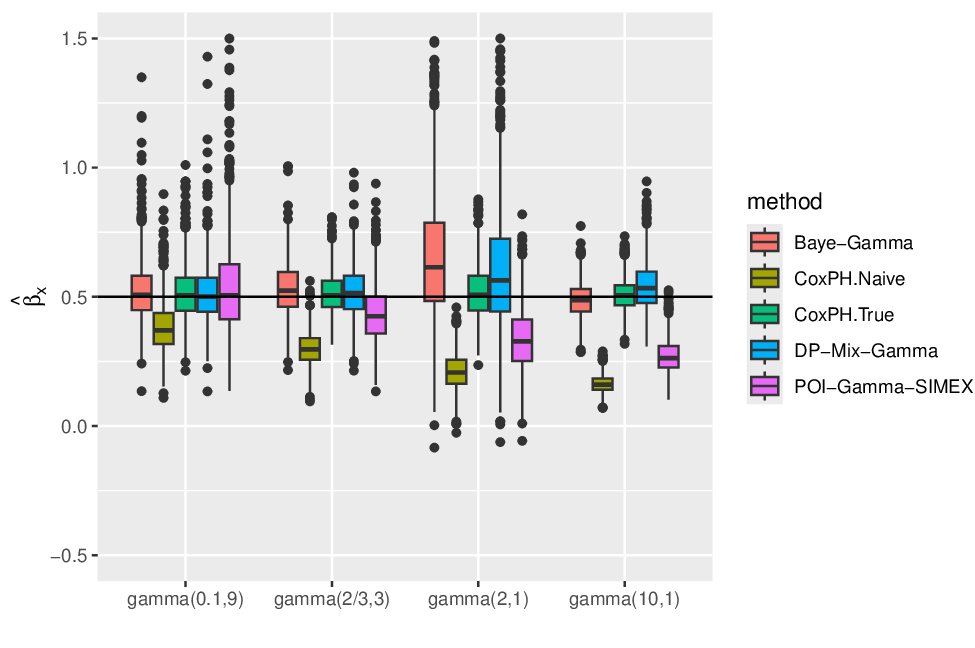}}
\caption{Boxplot of the $\beta_x$ estimates from each method across 1000 simulations with 20 \% censoring using a gamma distribution to simulate the true covariate.\label{f:fig1}}
\end{figure}
\begin{figure} [!htbp]
\centerline{\includegraphics[scale=0.7]{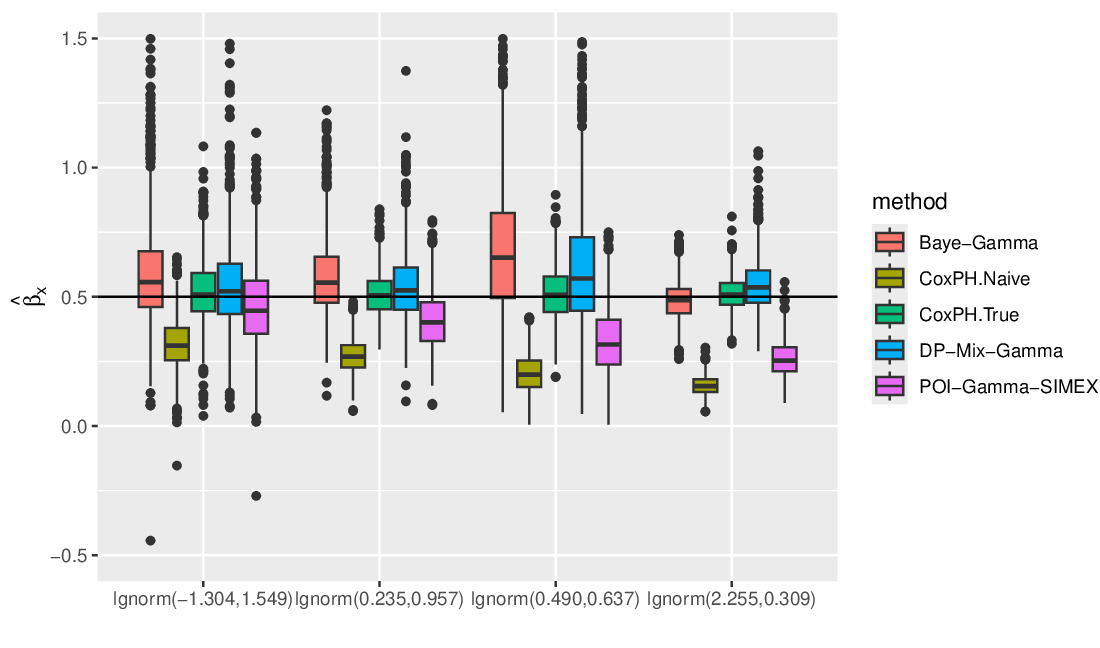}}
\caption{Boxplot of the $\beta_x$ estimates from each method across 1000 simulations with 20 \% censoring using a lognormal distribution to simulate the true covariate. \label{f:fig2}}
\end{figure}
\begin{figure} [!htbp]
\centerline{\includegraphics[scale=0.7]{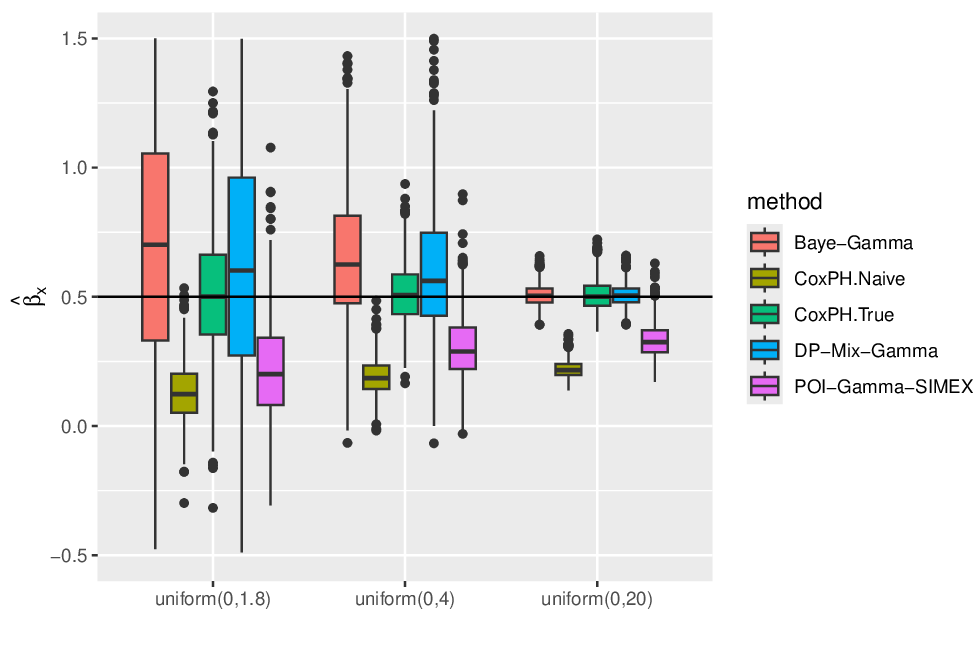}}
\caption{Boxplot of the $\beta_x$ estimates from each method across 1000 simulations with 20 percent censoring using the uniform distribution to simulate the true covariate. \label{f:fig3} }
\end{figure}
%%%Web Tables S1 - S4 in Web Appendix C--reference up to Appendix section
Bias, mean squared error (MSE), and their Monte Carlo (MC) standard errors are computed across all scenarios using 1,000 simulated datasets, employing the batch mean method to compute MC standard errors with ten batches \citep{Conway1963}. This is presented in Web Appendix C (Tables S1 - S4). The results of the Cox proportional hazard model estimation methods compare (i) regression with the true covariate; (ii) the surrogate; (iii) POI-Gamma-SIMEX error correction; and (iv) and (v) both the parametric and semiparametric Bayesian joint models. 

A total of eight simulation settings are conducted. Examining the MSE for $\beta_{x}$, as expected we find that the analysis using the true covariate has the lowest MSE in all cases. Following this, the semiparametric Bayesian joint model performs the best in four out of the eight settings. 
In the remaining four settings where it does not have the lowest MSE, it has the second lowest MSE compared with the other methods. The parametric Bayesian joint model has the lowest MSE in two settings and the SIMEX procedure in the other two. The performance of the SIMEX procedure is quite variable, and it actually performs worse than the naïve approach in one setting. Similarly, the performance of the parametric Bayesian joint model is also relatively more variable. Aside from the truth, the proposed semiparametric Bayesian joint model has the best performance in terms of MSE for $\beta_{x}$. 

Boxplots for estimates of $\beta_{x}$ for cases with $20$ percent random censoring are depicted in Figure~\ref{f:fig1}. These represent four of the eight simulation settings. Examining the location and variability of these boxplots it is clear, and as expected, that the regression with the true covariates and that with the naïve covariates yields the best and worst performance, respectively. Among the remaining three approaches, the bias associated with the SIMEX procedure is noticeably larger than the Bayesian approaches. Comparing the semiparametric and parametric Bayesian approaches, we note that the semiparametric approach generally performs as well or better than the correctly specified parametric joint model.

To assess the performance of these methods under misspecification of the underlying gamma distribution, we conduct an additional 14 sets of simulations representing seven settings with no censoring and 20\% censoring. In the first set of additional simulations, the true covariate is generated from a lognormal distribution with mean and variance matching those of the gamma distributions considered in the first set of simulations. In the next set of three simulations, the true covariate is drawn from a uniform distribution on the range of $(0,1.8)$, $(0,4)$ and $(0,20)$. 
%Tables S5 - S8 Web Appendix D
The results for the lognormal simulations are presented in Tables S5 - S8 in Web Appendix D and in Figure ~\ref{f:fig2}. From Figure 2 we see that the naïve approach tends to have the largest bias, followed by SIMEX, then followed by the parametric Bayesian joint model and finally the semiparametric Bayesian joint model which is closest to the method based on the true covariate in terms of bias. There are cases where the bias of SIMEX is substantial relative to the Bayesian approaches. On the other hand, SIMEX performs best according to its MSE which is as good or better than that of both Bayesian approaches in six out of eight settings. Under misspecification, the semiparametric Bayesian estimator has a lower MSE than its parametric counterpart in six out of eight settings, while its MSE is about twice as large in the other two.
%Table 9 - 11 in Web Appendix D
The results for the uniform simulations are presented in Tables S9 - S10 in Web Appendix D and in Figure ~\ref{f:fig3}. Similar conclusions tend to hold in this case with respect to bias though we notice that the variance of the Bayesian estimators increases markedly when the upper bound on the uniform distribution for the true density is 1.4 and 4. These simulations also indicate that the semiparametric Bayesian approach outperforms the parametric Bayesian approach under this form of misspecification.

Overall, it appears as though the semiparametric joint model exhibits the best performance generally and offers substantial robustness relative to the parametric joint model. The bias associated with the Poisson-Gamma adjusted SIMEX procedure can be substantial for $n=100$.

%We can get bold symbols using \verb+\bmath+, for example, $\bmath{\alpha}_i$.
%\citep{Arora2018}. \citep{Labidi2017}, \citep{Curiel2004,Curiel2008,Elkoshi2022} \citep{Leffers2009,Milne2009, Yang2024}
%\citep{Preston2013,Warfvinge2017,Liston2022}
\section{Application}
\label{s:app}
Epithelial ovarian cancer (EOC), though tenfold less prevalent than breast cancer, is the fifth-leading cause of cancer-related mortality among women \citep{Arora2018}. Despite therapeutic advancements over the past decade that have improved survival, outcomes remain suboptimal, with persistent challenges in survival rates and quality of life. High-grade serous carcinoma (HGSC), an aggressive subtype of epithelial ovarian cancer (EOC) that arises from the ovarian or fallopian tube epithelium \citep{Labidi2017}, motivates our investigation into the association between cell biomarkers and survival outcomes. As emphasized in the article and demonstrated in the simulation studies, it is important to account for the approximate nature of the discrete covariate. In this analysis we specifically investigate the association between survival and  CD3+CD8+FoxP3+ $\sim$E cell density.  While regulatory T cells (Tregs; FoxP3+) typically suppress antitumor immunity and correlate with poor prognosis \citep{Curiel2004,Curiel2008,Elkoshi2022}, their interplay with CD8+ T cells in ovarian cancer is complex. Prior studies associate CD8+FoxP3+ infiltration with improved survival in advanced EOC \citep{Leffers2009,Milne2009, Yang2024}, though their suppressive effects on B cells \citep{Preston2013,Warfvinge2017,Liston2022} and net impact in HGSC warrant further investigation.

This study leverages the Canadian Ovarian Cancer Research Consortium (COEUR), a biobank network comprising over 2,000 EOC tissue samples with clinical annotations, including 1,246 HGSC cases. Our cohort (n=716) focuses on HGSC patients with complete clinic and demographic data, treatment history, survival outcomes, and linked tissue microarray (TMA) data. 

Key clinical summary statistics are provided in Web Appendix E (Table S12). Kaplan–Meier curves and associated p-values from a log-rank test are presented in Web Appendix E (Figure S6). These suggest that prolonged survival is associated with optimal debulking and FIGO stage. Conversely, there does not appear to be an association between survival and age category. Finally, for CD3+CD8+FoxP3+ $\sim$E cell density, the situation is less clear. The Kaplan-Meier curves and their associated confidence intervals indicate some evidence that the presence of CD3+CD8+FoxP3+ $\sim$E is associated with improved survival. Care must be taken so as not to over-interpret these results, however, as the strata-specific comparisons have not been adjusted for the other variables such as FIGO stage. 

To better use the information on the biomarker, we avoid dichotomizing it and use the core density along with age, FIGO stage and debulking in a multivariable Cox regression analysis. 
We checked the assumed linear form for the biomarker of interest by fitting a standard multivariable Cox proportional hazard regression model and examining the martingale residual plot (Web Appendix F Figure S7). The diagnostic indicates that the assumed linear form is reasonable in this case. 

In addition to fitting the Cox model based on the naïve approach without adjustment for the surrogate, we apply the Poisson-gamma SIMEX procedure, and our semiparametric Bayesian joint model with 5 equal-length intervals for the piecewise constant baseline hazard and $K=5$. The results from the three approaches are summarized in Table~\ref{t:tab1}. The p-value and 95\% confidence intervals for the SIMEX estimator are computed using the standard normal approximation, following the estimation of its variance according to the procedure outlined in Section of 3.4 \citet{Yang2024}. For the Bayesian approach, we use $100,000$ MCMC samples thinned by $10$ after running $100,000$ burn-in iterations. We also applied the Dirichlet process joint model with 10 equal-spaced intervals for the baseline hazard and $K=10$ and the results are very similar to those reported Table~\ref{t:tab1}. 
%need to check 3.4 section number as we are going to cite from paper 1a

With regards to the hypothesis test of interest $H_{0}: \beta_{x} = 0$ versus $H_{1}: \beta_{x} \ne 0$ for the association between survival and CD3+CD8+FoxP3+ $\sim$E cell density, the naïve approach yields a p-value of $p=0.0326$, Poisson-Gamma adjusted SIMEX yields $p=0.047$ and the Bayesian approach yields a Bayes factor $B_{10} = 0.0068$. The naïve and SIMEX approaches are both statistically significant at a level of $\alpha = 0.05$, while the Bayes factor provides strong evidence in favor of the null hypothesis with a sample size of $n=716$ and $507$ deaths. The Bayesian test and the null hypothesis significance test results are thus contradictory when the latter are based on the standard significance threshold. Examining interval estimates, we obtain 95\% confidence intervals (CIs) for the associated hazard ratio as $(0.9784, 0.9991) $ from the Naïve model, $0.9859\, (0.9721, 0.9998)$ for the Poi-Gamma SIMEX model, with a 95 \%  highest posterior density interval $(0.9491, 1.0015)$ from the Bayesian model. 

\begin{figure}[t]
\centerline{\includegraphics[scale=1]{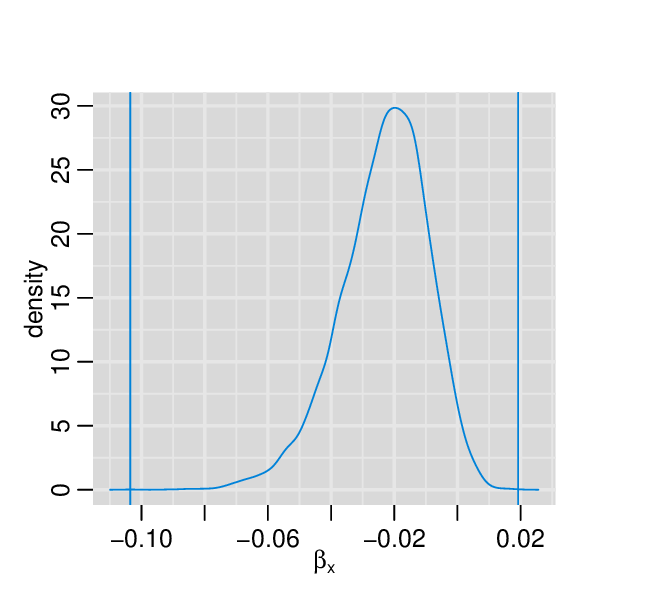}}
\caption{ Posterior density plot with 95\% highest posterior density intervals for the coefficient estimate of the biomarker CD3 + CD8 + FoxP3 + $\sim$ E cell density. \label{f:fig4}}
\end{figure}

The interval estimates all indicate an extremely small observed effect size for the association between survival and CD3+CD8+FoxP3+ $\sim$E cell density. In addition, the sample size is quite large, so that a standard level of statistical significance is attained with such an observed effect size. It is this small observed effect size and the large sample size that produces a contradiction between the null hypothesis significance test and the Bayes factor. The marginal posterior density of $\beta_{x}$ is plotted in Figure~\ref{f:fig4} with the lower and upper bounds of the 95\%  highest posterior density intervals delineated.

Since Bayes factors can be sensitive to prior distributions, we conduct a sensitivity analysis by refitting the model and computing the Bayes factor for five different prior distributions for $\beta_{x}$. All five priors are centered on the null hypothesis and four are Gaussian with variances $\sigma^{2} = 0.01, 1, 10, 100$. The fifth prior is a standard Cauchy distribution. These prior distributions, the resulting point and interval estimates, as well as the associated Bayes factors are presented in Table S13 in Web Appendix G. To be considered alongside these results, the posterior density associated with each prior is shown in Figure S8 in Web Appendix G. The posterior densities show a remarkable lack of sensitivity to the prior distribution for these data. Since the posterior density is essentially the same regardless of the prior, the Bayes factor varies only because of the height of the prior density at $\beta_{x} = 0$. This leads to considerable movement of the Bayes factors from as small as $BF_{10} = 0.0068$  (prior variance 100) to as large as $BF_{10} = 0.6624$ (prior variance 0.01). What is the same across all five cases in this reasonably broad range of priors is that the data provide evidence for the null hypothesis in moving from prior to posterior. Thus, it is clear that the data support the null hypothesis, indicating that CD3 + CD8 + FoxP3 + $\sim$ E cell density is not associated with survival in the COEUR cohort. 
%Bayes factors Web Table 13 and priors in Figure S8  in Web Appendix G

\begin{table}
\begin{center}
\caption{Cox PH regression with Poisson distributed biomarker CD3+CD8+FoxP3+ $\sim$E for the COEUR cohort.} \label{t:tab1}
\begin{tabular}{llllllll}
\Hline
Method & Parameters  & Coef.  & HR  & HR$_{Lower}$ & HR$_{Upper}$ & $p$-value\\ 
\hline
Naïve   & CD3+CD8+FoxP3+ $\sim$E  & -0.0114 & 0.9887 & 0.9784 & 0.9991 & 0.0326 \\     
& Age 60+ & 0.0999  & 1.1050  & 0.9158   & 1.3334  & 0.2974  \\   
& Debulking & & & & & \\
& \ \ Non-optimal & 0.5352  & 1.7079 & 1.3750     & 2.1214    & \textless{}0.0001        \\
& \ \ NA   & 0.5817  & 1.7891 & 1.3834    & 2.3137    & \textless{}0.0001    \\
& FIGO Stage (III/IV) & 0.8683  & 2.3828 & 1.7701    & 3.2077    & \textless{}0.0001        \\
\hline
POI-Gamma & CD3+CD8+FoxP3+ $\sim$E & -0.0142 & 0.9859 & 0.9721  & 0.9998  & 0.0471 \\ 
SIMEX   & Age 60+  & 0.1040   & 1.1096 & 0.9190  & 1.3399 & 0.2794 \\         
& Debulking & & & & & \\
& \ \ Non-optimal & 0.5345  & 1.7065 & 1.3733    & 2.1205    & \textless{}0.0001        \\
& \ \ NA & 0.5786  & 1.7836 & 1.3782    & 2.3082    & \textless{}0.0001        \\
& FIGO Stage (III/IV)         & 0.8670   & 2.3867 & 1.7724    & 3.2139    & \textless{}0.0001        \\
\hline
& & & & & & BF$_{10}$\\
DP-Mixture  & CD3+CD8+FoxP3+ $\sim$E& -0.0240  & 0.9763 & 0.9491 & 1.0015 &  0.0068 \\
& Age 60+ & 0.0994  & 1.1045 & 0.9158    & 1.3301    &    0.0162 \\
& Debulking & & & & & \\
& \ \ Non-optimal & 0.5216  & 1.6848 & 1.3423    & 2.0893    &    128    \\
& \ \ NA  & 0.5430   & 1.7212 & 1.3218    & 2.2324    &   10.7   \\
& FIGO Stage (III/IV)         & 0.8524  & 2.3454 & 1.7358    & 3.2038    &    2567 \\
\hline
\end{tabular}
\end{center}
\end{table}

\section{Discussion}
\label{s:discuss}
Motivated by cancer survival studies incorporating immune profiling  of tumor tissue using tissue microarrays, we have developed a semiparametric joint survival model for conditionally Poisson distributed biomarkers. Our proposed approach handles both the biomarker density distribution and baseline hazard function in a flexible manner while retaining computational tractability. Bayes factors are proposed for hypothesis testing. For the purpose of comparison, we also develop an adaptation of the SIMEX algorithm where simulations are based on a Poisson-gamma error structure.

An extensive set of simulation studies have been carried out to evaluate the proposed joint model. In a comparison of four approaches, the proposed model showed generally the best performance. It is also evident that modeling the biomarker density distribution with a Dirichlet process mixture yields a procedure with a greater degree of robustness compared to a fully parametric Bayesian joint model.

Our example examined the association between CD3+CD8+FoxP3+ $\sim$E density and ovarian cancer survival in an analysis of the COEUR cohort. Bayes factors computed over a reasonable range of prior distributions indicate the data provide evidence that  CD3+CD8+FoxP3+ $\sim$E density is not associated with survival. This is in contrast to what would be reported in a naïve analysis testing at level $\alpha = 0.05$.

While the developments in this paper have focused on cell density biomarkers derived from counts of the number of cells on each tissue core, the actual locations of each cell on each core are available as a spatial point pattern. This raises interesting opportunities for integrating the survival and tissue microarray data at a finer level, incorporating spatial point patterns into survival analysis. This may be useful in relaxing the underlying conditional homogeneous Poisson process assumption and integrating biomarker intensity functions as functional covariates in survival analysis.

\backmatter

%  This section is optional.  Here is where you will want to cite
%  grants, people who helped with the paper, etc.  But keep it short!

\section*{Acknowledgements}

We thank the patients who contributed tumor samples and clinical data to the COEUR cohort.
Nathoo and Lesperance acknowledge funding from NSERC (RGPIN$-04044-2020$, RGPIN$-07079-2020$) through the Discovery Grants Program.  Yang acknowledges funding from the University of Victoria and the Pacific Leaders Scholarship Program. Nelson acknowledges the BC Cancer Foundation, CIHR (PJT$-427647$),  and Terry Fox Research Institute (TFRI$1060$)for funding. Lum acknowledges funding from CIHR (PJT$-192015$), Terry Fox Research Institute, and Lotte \& John Hecht Memorial Foundation (TFRI$1125$). 

\vspace*{-8pt}

%  If your paper refers to supplementary web material, then you MUST
%  include this section!!  See Instructions for Authors at the journal
%  website http://www.biometrics.tibs.org

\section*{Supplementary Materials}
Web Appendices containing figures and tables referenced in Section~\ref{s:method} to ~\ref{s:app}, are available with this paper at the Biometrics website on Oxford Academic. R programs for data analysis are available at
GitHub (https://github.com/Aijunyan/DP-mix-gamma-coxPH) \vspace*{-8pt}

%%%two covariates simulation studies moved to the supplmentary section

%  Here, we create the bibliographic entries manually, following the
%  journal style.  If you use this method or use natbib, PLEASE PAY
%  CAREFUL ATTENTION TO THE BIBLIOGRAPHIC STYLE IN A RECENT ISSUE OF
%  THE JOURNAL AND FOLLOW IT!  Failure to follow stylistic conventions
%  just lengthens the time spend copyediting your paper and hence its
%  position in the publication queue should it be accepted.

%  We greatly prefer that you incorporate the references for your
%  article into the body of the article as we have done here 
%  (you can use natbib or not as you choose) than use BiBTeX,
%  so that your article is self-contained in one file.
%  If you do use BiBTeX, please use the .bst file that comes with 
%  the distribution. %\bibliographystyle{biom} \bibliography{mybiom}

\begin{thebibliography}{}
\bibitem [\protect\citeauthoryear {Antoniak}{1974}]{Antoniak1974}
Antoniak, C. E. (1974). Mixtures of Dirichlet processes with applications to Bayesian nonparametric problems. {\it The annals of statistics}, 1152-1174.
\bibitem [\protect\citeauthoryear {Arora et al.}{2018}] {Arora2018}
Arora, N., Talhouk, A., McAlpine, J., Law, M. \& Hanley, G. (2018). Long-term mortality among women with epithelial ovarian cancer: a population-based study in British Columbia, Canada. {\it BMC Cancer}. \textbf{18} pp. 1039, https://doi.org/10.1186/s12885-018-4970-9.
\bibitem [\protect\citeauthoryear {Augustin} {2004}] {Augustin2004} 
Augustin, T. (2004). An Exact Corrected Log-Likelihood Function for Cox’s Proportional Hazards Model under Measurement Error and Some Extensions. {\it Scandinavian Journal Of Statistics}. \textbf{31}, 43-50, https://www.jstor.org/stable/4616810.
\bibitem [\protect\citeauthoryear {Carroll et al.} {1999}] {Carroll1999}
Carroll, R., Delaigle, A. \& Hall, P. (1999). Nonparametric Regression in the Presence of Measurement Error. \textit{Biometrika}. \textbf{86}, http://www.jstor.org/stable/2673653.
\bibitem  [\protect\citeauthoryear {Carrol et al.} {2006}]{Carroll2006}
 Carroll, R., Ruppert, D., Stefanski, L. \& Crainiceanu, C. (2006) \textit{Measurement Error in Nonlinear Models: A Modern Perspective}. Chapman and Hall, London.
\bibitem [\protect\citeauthoryear {Carroll et al.} {2009}] {Carroll2009}
 Carroll, R., Delaigle , A. \& Hall, P. (2009). Nonparametric Prediction in Measurement Error Models. \textit{J Am Stat Assoc}. \textbf{104}, 10.1198/jasa.2009.tm07543.
\bibitem [\protect\citeauthoryear {Cao \& Wong} {2022}] {Cao2022}
Cao, Z. \& Wong, M. (2022). Approximate profile likelihood estimation for Cox regression with covariate measurement error. \textit{Stat Med.}. \textbf{41}, 910-931, https://doi: 10.1002/sim.9324. 
\bibitem  [\protect\citeauthoryear {Chen \& Yi} {2020}]{Chen2020}
 Chen, L., \& Yi, G. Y. (2020). Graphical proportional hazards measurement error models. {\it Biometrics}, \textbf{76(2)}, 463-474.
\bibitem  [\protect\citeauthoryear {Chen \& Yi} {2024}]{Chen2024}
Chen, L., \& Yi, G. Y. (2024). SIMEXBoost: Simulation extrapolation for high-dimensional survival data with measurement error. {\it Journal of Computational and Graphical Statistics}, \textbf{33(1)}, 1-15.
\bibitem  [\protect\citeauthoryear {Conway} {1963}]{Conway1963}
Conway, R. W. (1963). Some tactical problems in digital simulation. {\it Management science}, \textbf{10(1)}, 47-61.
\bibitem  [\protect\citeauthoryear {Cook and Stefanski} {1994}]{Cook1994} Cook, J. \& Stefanski, L. (1994). 
Simulation-Extrapolation Estimation in Parametric Measurement Error Models. {\it J Am Stat Assoc}. \textbf{89}, https://doi.org/10.1080/01621459.1994.10476871.
\bibitem  [\protect\citeauthoryear {Cox} {1972}] {Cox1972} Cox, D. R. (1972). Regression models and life tables (with discussion). {\it Journal of the Royal Statistical Society, Series B}
\textbf{34,} 187--200.
\bibitem [\protect\citeauthoryear {Christensen et al.}{2010}] {christensen2010}
Christensen, R., Johnson, W., Branscum, A., Hanson, T.E. (2010). Bayesian ideas and data analysis: an introduction for scientists and statisticians. CRC press.
\bibitem [\protect\citeauthoryear {Curiel et al.}{2004}] {Curiel2004}
Curiel, T., Coukos, G., Zou, L., Alvarez, X., Cheng , P., Mottram, P., Evdemon-Hogan, M., Conejo-Garcia, J., Zhang, L., Burow, Zhu, Y., Wei, S., Kryczek, I., Daniel, B., Gordon, A., Myers, L., Lackner, A., Disis, M., Knutson, K., Chen, L. \& Zou, W. (2004). Specific recruitment of regulatory T cells in ovarian carcinoma fosters immune privilege and predicts reduced survival. {\it Nat Med.}. \textbf{10} , https://doi: 10.1038/nm1093
\bibitem [\protect\citeauthoryear {Curiel}{2008}]{Curiel2008}Curiel, T. (2008). Regulatory T cells and treatment of cancer. {\it Current Opinion In Immunology}. \textbf{20}, 241-246, https://www.sciencedirect.com/science/article/pii/S0952791508000502.
\bibitem  [\protect\citeauthoryear {Devanarayan \& Stefanski}{2002}]{Devanarayan2002}
Devanarayan, V., \& Stefanski, L. A. (2002). Empirical simulation extrapolation for measurement error models with replicate measurements. {\it Statistics \& Probability Letters}, \textbf{59(3)}, 219-225.
\bibitem  [\protect\citeauthoryear {Dunson \& Xing} {2009}]{Dunson2009}
Dunson, D. B., \& Xing, C. (2009). Nonparametric Bayes modeling of multivariate categorical data. {\it Journal of the American Statistical Association}, \textbf{104(487)}, 1042-1051.
\bibitem [\protect\citeauthoryear {Elkoshi} {2022}] {Elkoshi2022} Elkoshi, Z. (2022). On the Prognostic Power of Tumor Infiltrating Lymphocytes – A Critical Commentary. {\it Front. Immunol}. \textbf{13}, https:doi: 10.3389/fimmu.2022.892543.
\bibitem [\protect\citeauthoryear {Ferguson} {1973}] {Ferg1973}
Ferguson, T. S. (1973). A Bayesian analysis of some nonparametric problems.{\it Annals of Statistics}, \textbf{1(2)}, 209–230, https://doi.org/10.1214/aos/1176344136. 
\bibitem [\protect\citeauthoryear {Fuller} {1987}] {Fuller1987}
 Fuller, W. (1987). \textit{Measurement error models}. John Wiley \& Sons, Inc., USA.  
\bibitem [\protect\citeauthoryear {Gelman et al.} {2013}] {Gelman2013}
Gelman, A., Carlin, J. B., Stern, H. S., Dunson, D. B., Vehtari, A., \& Rubin, D. B. (2013) \textit {Bayesian Data Analysis (3rd ed.)}. Chapman and Hall/CRC.
\bibitem [\protect\citeauthoryear {Gustafson} {2003}]{Gustafson2003}
Gustafson, P. (2003). {\it Measurement error and misclassification in statistics and epidemiology: Impacts and Bayesian adjustments}. CRC Press.
\bibitem [\protect\citeauthoryear {Greene \& Cai} {2004}]{Greene2004}
Greene, W. \& Cai, J. Measurement error in covariates in the marginal hazards model for multivariate failure time data. {\it Biometrics}.\textbf{60}, 987-996.
\bibitem  [\protect\citeauthoryear {Havasi et al.} {2023}] {Havasi2023} 
Havasi, A., Cainap, S.S., Havasi, A.T., \& Cainap, C. (2023). Ovarian Cancer—Insights into Platinum Resistance and Overcoming It. {\it Medicina.}, \textbf{59(3)}, 544. https://doi.org/10.3390/medicina59030544.
\bibitem  [\protect\citeauthoryear {He et al.} {2007}] {He2007}
He, W., Yi, G. Y., \& Xiong, J. (2007). Accelerated failure time models with covariates subject to measurement error. {\it Stat Med}. \textbf{26(26)}, 4817-4832.https://doi.org/10.1002/sim.2892
\bibitem  [\protect\citeauthoryear {He et al.} {2012}] {He2012}
He, W., Yi, G. \& Xiong, J.  (2012). SIMEX R Package for Accelerated Failure Time Models with Covariate Measurement Error. {\it Journal of Statistical Software, Code Snippets}, \textbf{46(1)}, 1–14. https://doi.org/10.18637/jss.v046.c01
\bibitem  [\protect\citeauthoryear {Ibrahim et al.} {2001}] {Ibrahim2001} 
Ibrahim, J. G., Chen, M.H., \& Sinha, D. (2001) Bayesian Survival Analysis. \textit{Springer Series in Statistics, First Edition},  Springer, New York. http://dx.doi.org/10.1007/978-1-4757-3447-8.
\bibitem [\protect\citeauthoryear {Kass \& Raftery} {1995}] {KassRaftery1995}
Kass, R. E., \& Raftery, A. E. (1995). Bayes factors. {\it Journal of the American Statistical Association}, \textbf{90 (430)}, 773-795. https://doi.org/10.1080/01621459.1995.10476572
\bibitem  [\protect\citeauthoryear {Kong \& Gu} {1999}] {Kong1999}
Kong, F. H., \& Gu, M. (1999). Consistent estimation in Cox Proportional hazards model with covariate measurement errors. {\it Statistica Sinica}, \textbf{9(4)}, 953–969. http://www.jstor.org/stable/24306629
\bibitem [\protect\citeauthoryear {Labidi-Galy et al.} {2017}] {Labidi2017} Labidi-Galy, S., Papp, E., Hallberg, D., Niknafs, N., Adleff, V., Noe, M., Bhattacharya, R., Novak, M., Jones, S., Phallen , J., Hruban, C., Hirsch, M., Lin, D., Schwartz, L., Maire, C., Tille, J., Bowden, M., Ayhan, Wood, L., Scharpf, R., Kurman, T., Shih, I., Karchi, R. \& Drapkin, V. (2017). High grade serous ovarian carcinomas originate in the fallopian tube. {\it Nat Commun}. \textbf{8}, https://doi: 10.1038/s41467-017-00962-1.
\bibitem [\protect\citeauthoryear {Lee et al.} {2008}] {DP2008}
Lee, S. Y., Lu, B.,\& Song, X. Y. (2008). Semiparametric Bayesian analysis of structural equation models with fixed covariates. {\it Statistics in Medicine}, \textbf{27}, 2341–2360.
\bibitem [\protect\citeauthoryear {Leffers et al.}{2009}] {Leffers2009}Leffers, N., Gooden, M., Jong, R., Hoogeboom , B., Hoor, K., Hollema, H. \& et al. (2009). Prognostic Significance of Tumor-Infiltrating T-Lymphocytes in Primary and Metastatic Lesions of Advanced Stage Ovarian Cancer. {\it Cancer Immunol Immun}. \textbf{58}, https://doi: 10.1007/s00262-008-0583-5.
\bibitem [\protect\citeauthoryear {Li et al.} {2004}]{Li2004poi}
Li, L, Palta, M, Shao, J (2004). A measurement error model with a Poisson distributed surrogate.{\it Stat Med.} \textbf{23(16)}, 2527-2536. https://doi: 10.1002/sim.1838.
\bibitem [\protect\citeauthoryear {Li \& Ryan} {2006}]{Li2006}
Li, Y. \& Ryan, L. (2006), Inference on Survival Data with Covariate Measurement Error – An Imputation-based Approach. {\it Scandinavian Journal of Statistics}, \textbf{33:} 169-190. https://doi.org/10.1111/j.1467-9469.2006.00460.x.
\bibitem [\protect\citeauthoryear  {Lin \& Carroll} {2000}] {Lin2000}
Lin, X. \& Carroll, R.(2000). Nonparametric function estimation for clustered data when the predictor is measured without/with error. {\it Am Stat Assoc}. \textbf{95} pp. 520-534.http://www.jstor.com/stable/2333251.
\bibitem [\protect\citeauthoryear  {Lindley} {1957}] {Lindley1957}
Lindley, D. V. (1957). A statistical paradox. {\it Biometrika}. \textbf{44} pp. 187-192.
\bibitem [\protect\citeauthoryear {Liston \& Aloulou} {2022}] {Liston2022} Liston, A. \& Aloulou, M.(2022). A fresh look at a neglected regulatory lineage: CD8+Foxp3+ Regulatory T cells. {\it Immunology Letters}. \textbf{247}, https://doi.org/10.1016/j.imlet.2022.05.004.
\bibitem [\protect\citeauthoryear {Lunn et al.} {2012}] {Bugs2012}
Lunn, D., Jackson, C., Best, N., Thomas, A., \& Spiegelhalter, D. (2012) The BUGS Book: A Practical Introduction to Bayesian Analysis (1st ed.). Chapman and Hall/CRC.
\bibitem [\protect\citeauthoryear {Ma and Chen} {2021}] {DP2021}
Ma, Z., and Chen, G. (2021). Bayesian joint analysis using a semiparametric latent variable model with non-ignorable missing covariates for CHNS data. {\it Statistical Modeling}, \textbf{21(4)}, 313–331. https://doi.org/10.1177/1471082X19896688.
\bibitem [\protect\citeauthoryear {Milne et al.}{2009}]{Milne2009}Milne, K., Köbel, M., Kalloger, S., Barnes , R., Gao, D., Gilks, C., Watson, P. \& Nelson, B.(2009). Systematic Analysis of Immune Infiltrates in High-Grade Serous Ovarian Cancer Reveals CD20, FoxP3 and TIA-1 as Positive Prognostic Factors. {\it PLoS ONE}. \textbf{4}, https://doi.org/10.1371/journal.pone.0006412.
\bibitem [\protect\citeauthoryear {Müller et al.} {2015}]{Muller2015}
Müller, P., Quintana, F. A., Jara, A., \& Hanson, T. (2015). Bayesian nonparametric data analysis (1st ed. 2015.). Springer International Publishing. https://doi.org/10.1007/978-3-319-18968-0.
\bibitem [\protect\citeauthoryear {Nakamura}{1992}] {Nakamura1992}
Nakamura, T. (1992). Proportional hazards model with covariates subject to measurement error. {\it Biometrics}, \textbf{48(3)}, 829-838. https://doi.org/10.2307/2532348.
\bibitem [\protect\citeauthoryear {Ohlssen et al.}{2007}] {Ohlssen2007}
Ohlssen, D. I., Sharples, L. D., \& Spiegelhalter, D. J. (2007).Flexible random‐effects models using Bayesian semi‐parametric models: applications to institutional comparisons. {\it Statistics in medicine}, \textbf{26(9)}, 2088-2112. https://doi.org/10.1002/sim.2666.
\bibitem  [\protect\citeauthoryear {Pan} {2022}]{Pan2022} Pan, A. (2022).
Bayesian Nonparametric Approaches for Survival Data with Covariate Measurement Error. {\it Doctoral Dissertation, University of Georgia}.
https://esploro.libs.uga.edu/esploro/outputs/doctoral/\\
Bayesian-Nonparametric-Approaches-for-Survival-Data/9949450629202959\#file-0.
\bibitem [\protect\citeauthoryear  {Parveen et al.} {2021}] {Parveen2021} 
Parveen, N., Moodie, E. \& Brenner, B. (2021). The non-zero mean SIMEX: Improving estimation in the face of measurement error. {\it Observational Studies}. \textbf{1} pp. 90-123. 
\bibitem [\protect\citeauthoryear  {Plummer, M.} {2019}] {Plummer2019} 
Plummer, M. (2019). rjags: Bayesian Graphical Models using MCMC.R package version 4-10. https://CRAN.R-project.org/package=rjags. 
\bibitem [\protect\citeauthoryear {Prentice}{1982}]{Prentice1982}
Prentice, R. (1982). Covariate measurement errors and parameter estimation in a failure time regression model. {\it Biometrika}. \textbf{69}, 331-342.
\bibitem [\protect\citeauthoryear {Preston et al.} {2013}] {Preston2013} Preston, C., Maurer, M., Oberg, A., Visscher, D., Kalli, K., Hartmann, L., Goode, E. \& Knutson, K. (2013). The ratios of CD8+ T cells to CD4+CD25+ FOXP3+ and FOXP3- T cells correlate with poor clinical outcome in human serous ovarian cancer. {\it PLoS One}. \textbf{8}, https://10.1371/journal.pone.0080063.
\bibitem [\protect\citeauthoryear {Richardson \& Gilks} {1993}] {Richardson1993}
Richardson, S. \& Gilks , W. (1993). A Bayesian approach to measurement error problems in epidemiology using conditional independence models. {\it Am J Epidemiol}. \textbf{138}. https://doi:10.1093/oxfordjournals.aje.a116875.
\bibitem [\protect\citeauthoryear {Sarkar}  {2014}] {Sarkar2014}
Sarkar, A. (2014). Bayesian Semiparametric Density Deconvolution and Regression in the Presence of Measurement Errors. {\it Doctoral Dissertation, Texas A \& M University}. https://hdl.handle.net/1969.1/153327.
\bibitem [\protect\citeauthoryear {Sethuraman} {1994}] {DP1994}
Sethuraman, J. (1994). A constructive definition of Dirichlet priors. {\it Statistica sinica}, \textbf{}, 639–650.
\bibitem [\protect\citeauthoryear {Sinha \& Wang} {2017}] {Sinha2017}
Sinha, S. \& Wang, S. Semiparametric Bayesian analysis of censored linear regression with errors-in-covariates. {\it Statistical Methods In Medical Research}. \textbf{26}, 1389-1415. doi:10.1177/0962280215580668.
\bibitem [\protect\citeauthoryear {Spiegelman et al.} {1997}] {Spiegelman1997} 
Spiegelman, D., McDermott, A. \& Rosner, B. (1997). Regression calibration method for correcting measurement-error bias in nutritional epidemiology. {\it Am J Clin Nutr}. \textbf{65}, https://doi:10.1093/ajcn/65.4.1179S.
\bibitem [\protect\citeauthoryear {Tang et al.} {2017}] {DP2017}
Tang, A. M., Zhao, X.,\& Tang, N. S. (2017). Bayesian variable selection and estimation in semiparametric joint models of multivariate longitudinal and survival data. {\it Biometrical Journal}, \textbf{59(1)}, 57–78.
\bibitem [\protect\citeauthoryear {Tang} {2022}] {Tang2022}
Tang, Z. (2022). Bayesian accelerated failure time models with covariate measurement error: A flexible approach. {\it Statistical Methods in Medical Research}, \textbf{31(5)}, 858-874.
\bibitem [\protect\citeauthoryear {Wagenmakers et al.} {2010}]{Wagenmakers2010}
Wagenmakers, E. J., Lodewyckx, T., Kuriyal, H., \& Grasman, R. (2010). Bayesian hypothesis testing for psychologists: A tutorial on the Savage–Dickey method. {\it Cognitive psychology}, \textbf{60(3)}, 158-189.
\bibitem [\protect\citeauthoryear {Warfvinge et al}{2017}] {Warfvinge2017}Warfvinge, C., Lundgren, S., Elebro, J., Margareta Heby, M., Krzyzanowska, A., Bjartell, A., Nodin, B., Jakob Eberhard, J., Leandersson, K. \& Jirstrom, K. (2017). The prognostic impact of CD3, CD8, FoxP3, and IL17 tumor-infiltrating immune cells in periampullary cancer differs by morphological type and adjuvant chemotherapy. {\it Clinical Oncology}. \textbf{35}, https://doi.org/10.1200/JCO.2017.35.7-suppl.53.
\bibitem [\protect\citeauthoryear {Wang et al.} {1997}] {Wang1997_RC}
Wang, C., Hsu, L., Feng, Z. \& Prentice, R. (1997).
Regression calibration in failure time regression. {\it Biometrics}. \textbf{53} pp. 131-145.
\bibitem [\protect\citeauthoryear {Wang \& Wang} {2008}] {Wang2008}
Wang, L. \& Wang, X. (2008). A new method for survival analysis with covariates subject to measurement error. {\it Journal Of The Royal Statistics Society:Series B}. \textbf{70}, 519-530.
\bibitem [\protect\citeauthoryear {Wei et al.} {2022}] {Wei2022}
Wei Z., Yang A., Rocha L., Miranda M. F., Nathoo F. S. A Review of Bayesian Hypothesis Testing and Its Practical Implementations. {\it Entropy (Basel)}.\textbf{24(2)}:161, https://doi: 10.3390/e24020161.
\bibitem [\protect\citeauthoryear {Lederer and Seibold} {2019}] {simex2019} Lederer, W. and Seibold H. (2019). simex: SIMEX- And MCSIMEX-Algorithm for Measurement Error Models, R package version 1.8, https://CRAN.R-project.org/package=simex.
\bibitem [\protect\citeauthoryear {Yan \& Yi} {2015}] {Yan2015} Yan, Y. \& Yi, G.(2015). A corrected profile likelihood method for survival data with covariate measurement error under the Cox model. {\it Can J Stat}. \textbf{43} pp. 454-480.
\bibitem [\protect\citeauthoryear{Yang et~al.}{2024}]{Yang2024} 
Yang, A., Hamilton, P. T., Nelson, B. H., Lum, J. J., Lesperance, M. \& Nathoo, F. S., (2024). POI-SIMEX for Conditionally Poisson Distributed Biomarkers from Tissue Histology. {\it arXiv 2409.14256}, https://arxiv.org/abs/2409.14256.
\bibitem [\protect\citeauthoryear{Yang et~al.}{2025}]{Yang2025} 
Yang, A., Lesperance, M. \& Nathoo, F. S., (2025). Strong Consistency of the SIMEX Estimator in Linear Regression with a Conditionally Poisson Covariate. {\it arXiv 2509.04709}, https://arxiv.org/abs/2509.04709.
%https://doi.org/10.48550/arXiv.2509.04709
\bibitem[\protect\citeauthoryear {Yi \& Lawless} {2007}] {Yi2007}
Yi, G. \& Lawless, J.(2007) A corrected likelihood method for the proportional hazards model with covariates subject to measurement error. {\it J Stat Plan Infer}. \textbf{137} pp. 1816-1828.
\bibitem[\protect\citeauthoryear {Yi et al.} {2021}]{Yi2021}
Yi, G., Delaigle, A. \& Gustafson, P. (2021) {\it Handbook of Measurement Error Models (1st ed.)}. Chapman and Hall/CRC, New York, https://doi.org/10.1201/9781315101279.
\bibitem [\protect\citeauthoryear {Zhang et al.}{2014}] {Zhang2014}
Zhang, J., He, W. \& H, L. (2014). A semiparametric approach for accelerated failure time models with covariates subject to measurement error. {\it Commun Stat Simul Comput}. \textbf{43}, 329-341.
\end{thebibliography}

%%%if need can add late
%\appendix

%  To get the journal style of heading for an appendix, mimic the following.

%\section{}
%\subsection{Title of appendix}

%Put your short appendix here.  Remember, longer appendices are
%possible when presented as Supplementary Web Material.  Please 
%review and follow the journal policy for this material, available
%under Instructions for Authors at \texttt{http://www.biometrics.tibs.org}.

\label{lastpage}
\end{document}

%  This will produce the submission and review information that appears
%  right after the reference section.  Of course, it will be unknown when
%  you submit your paper, so you can either leave this out or put in 
%  sample dates (these will have no effect on the fate of your paper in the
%  review process!)
\date{{\it Received xxx} 2025. {\it Revised xxx} 2025.  {\it Accepted xxx} 2025.}
\maketitle
%\clearpage
\beginsupplement

\begin{figure*}
\section*{Web Appendix A}
\centerline{\includegraphics[scale=1]{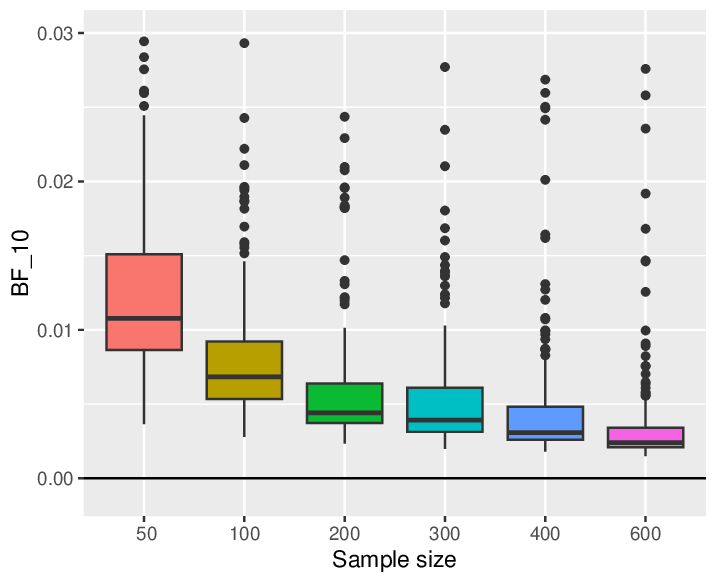}}
\vspace{-0.2cm}
\caption{Sampling distribution of the Bayes factor under $H_0$ for increasing sample sizes.}
\vspace*{-3pt}
\label{f:sfig1}
\vspace{0.5cm}
%\end{figure}
%\begin{figure}
\centerline{\includegraphics[scale=1]{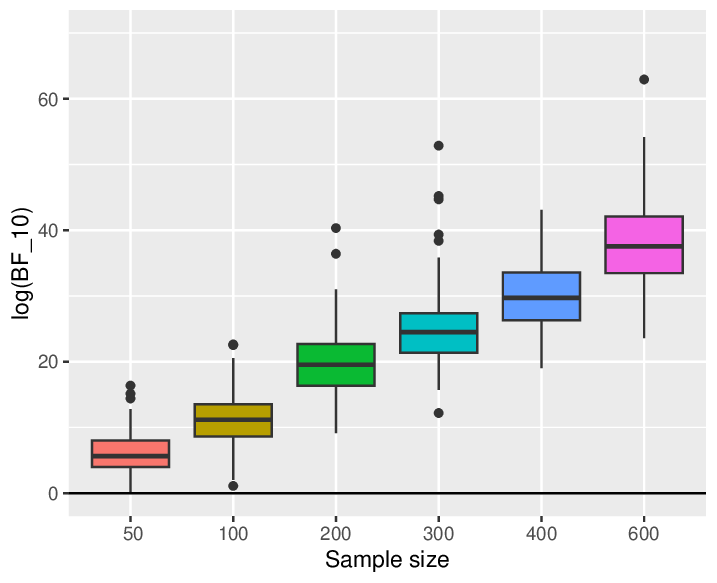}}
\vspace{-0.2cm}
\caption{Sampling distribution of the Bayes factor under $H_1$ for increasing sample sizes.}
\label{f:sfig2}
\end{figure*}
\begin{figure*}[b]
\section*{Web Appendix B}
\centerline{\includegraphics[scale=0.6]{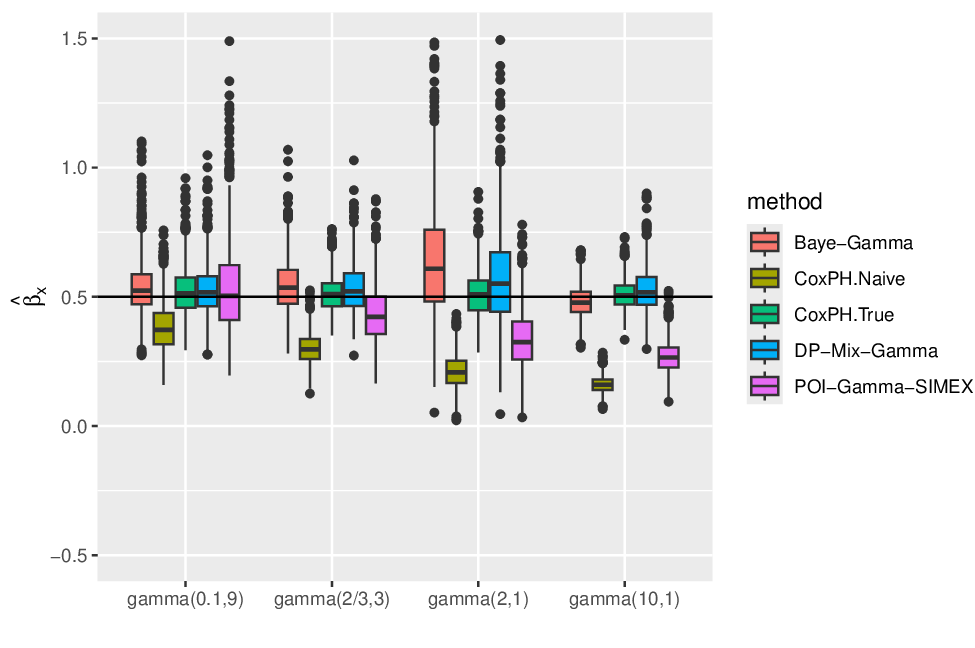}}
\vspace{-0.2cm}
\caption{Boxplot of the $\beta_x$ estimates from each method across 1000 simulations without censoring using the gamma distribution to simulate true covariate. }
\vspace*{-3pt}
\label{f:sfig3}
\vspace{0.5cm}
%\end{figure}
%\begin{figure}
\centerline{\includegraphics[scale=0.6]{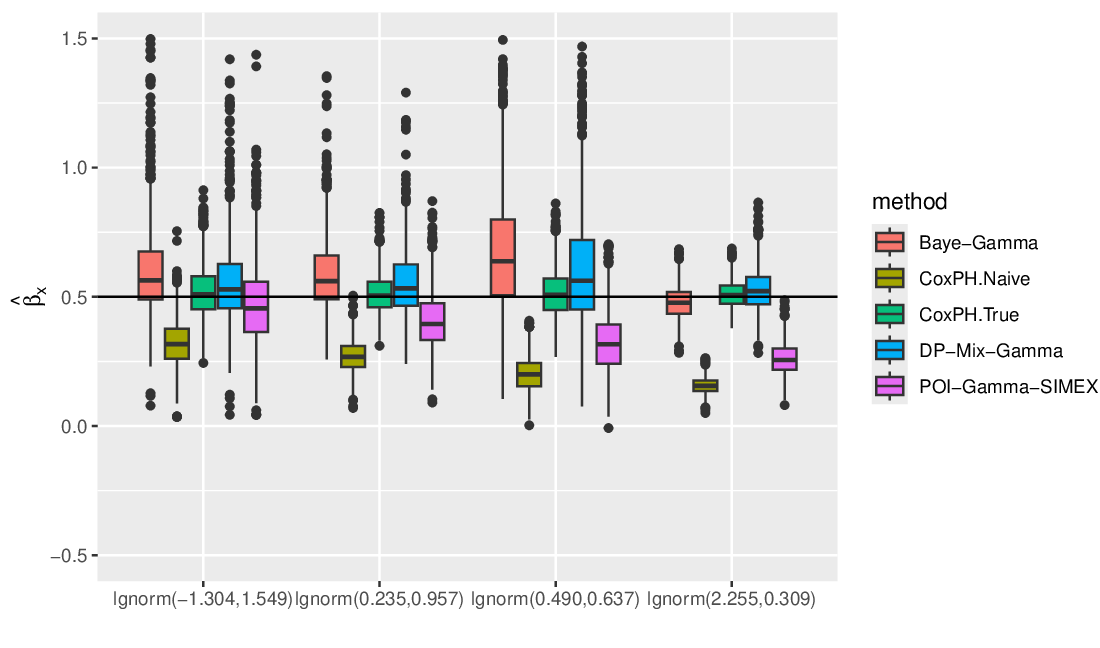}}
\vspace{-0.2cm}
\caption{Boxplot of the $\beta_x$ estimates from each method across 1000 simulations without censoring using the lognormal distribution to simulate the true covariate.}
\vspace*{-3pt}
\label{f:sfig4}
\vspace{0.5cm}
%\end{figure}
%\begin{figure}
\centerline{\includegraphics[scale=0.6]{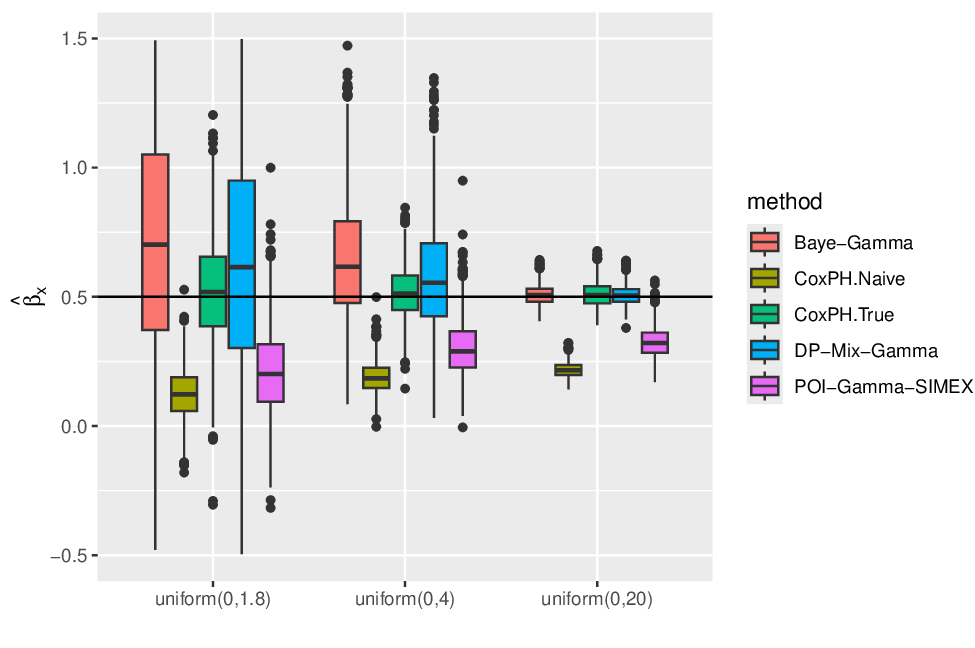}}
\vspace{-0.2cm}
\caption{Boxplot of the $\beta_x$ estimates from each method across 1000 simulations without censoring using uniform to simulate true covariate.}
\label{f:sfig5}
\vspace{-0.5cm}
\end{figure*}
%\begin{figure*}
%\section*{Web Appendix C}
%\centerline{\includegraphics[scale=0.80]{YangA_KM_plot.eps}}
%\caption{Kaplan-Meier survival curves with pointwise 95\% confidence interval band, log-rank test p-value, and median survival time (dash line) by strata.}
%\label{f:sfig6}
%\end{figure*}

%\begin{figure*}
%\centerline{\includegraphics[scale=0.80]{HGSC_bio_betax.eps}}
%\caption{HGSC coefficient estimate density plot of the biomarker}
%\label{f:sfig8}
%\end{figure*}
\clearpage
\begin{sidewaystable}[h]
\section*{Web Appendix C}
 \centering
 \def\~{\hphantom{0}} 
\caption{Simulation results for scenario 1: $X_i \sim \text{Gamma}(0.1,9)$, $W_i \lvert X_i \sim \text{Poisson} (X_i)$, without censoring and 20\% censoring. }
\label{t:tab1}
\tabcolsep=0pt%%
\begin{tabular*}{\textwidth}{@{\extracolsep{\fill}}llcccccc@{\extracolsep{\fill}}}
%\toprule%
\hline
%& & \multicolumn{3}{@{}c@{}}{without censoring$^{1}$} & \multicolumn{3}{@{}c@{}}{20\% censoring$^{2}$} \\
& & \multicolumn{3}{@{Mean Estimate}c@{}}{Without Censoring$^{1}$} & \multicolumn{3}{@{Mean Estimate}c@{}}{20\% Random Censoring} \\
\cline{3-5}\cline{6-8}%
Parameter & Methods & Estimator ($\hat{\beta}_x$)  & MSE(MCSE) & Bias (MCSE) &  Estimator ($\hat{\beta}_x$) & MSE(MCSE) & Bias (MCSE) \\
%\midrule
\hline
$\beta_x=0.5$ &Baye-Gamma  & 0.5363    & 0.0125(0.0008) & 0.0363(0.0013)  & 0.5273   & 0.0276(0.0095) & 0.0273(0.0044)  \\
& CoxPH.True         & 0.5216    & 0.0090(0.0005) & 0.0216(0.0024)  & 0.5171   & 0.0111(0.0009) & 0.0171(0.0045)  \\
& CoxPH.Naive         & 0.3813    & 0.0226(0.0007) & -0.1187(0.0017) & 0.3824   & 0.0241(0.0008) & -0.1176(0.0034) \\
& DP-mix-Gamma & 0.5287   & 0.0105(0.0006) & 0.0287(0.0024)  & 0.5176 & 0.0175(0.0028)	& 0.0176(0.0045) \\
& POI-Gamma-SIMEX & 0.5333    & 0.0329(0.0019) & 0.0333(0.0029)  & 0.5398   & 0.0376(0.0024) & 0.0398(0.0061)  \\
%\botrule
\hline
Parameter & Methods & Estimator ($\hat{\beta}_z$)  & MSE(MCSE) & Bias (MCSE) &  Estimator ($\hat{\beta}_z$) & MSE(MCSE) & Bias (MCSE) \\
%\midrule\
\hline
$\beta_z=0.1$ & Baye-Gamma  & 0.1025    & 0.0123(0.0005) & 0.0025(0.0047)  & 0.1081   & 0.0159(0.0007) & 0.0081(0.0049)  \\
&CoxPH.True & 0.1031    & 0.0113(0.0004) & 0.0031(0.0040)  & 0.1087   & 0.0148(0.0007) & 0.0087(0.0047)  \\
&CoxPH.Naive        & 0.0984    & 0.0119(0.0004) & -0.0016(0.0042) & 0.1054   & 0.0153(0.0007) & 0.0054(0.0050)  \\
& DP-mix-Gamma & 0.1017    & 0.0120(0.0004) & 0.0017(0.0040)  & 0.1061 & 0.0156(0.0005)& 0.0061(0.0058)\\
&POI-Gamma-SIMEX & 0.1023    & 0.0148(0.0005) & 0.0023(0.0042)  & 0.1095   & 0.0183(0.0010) & 0.0095(0.0055) \\
%\botrule
\hline
\end{tabular*}
\vspace*{0.8cm}
\begin{tablenotes}%
\item Note: Sample size n=100.% and Bayesian Gamma Poisson model failed to return the result.  
\item[$^{1}$] In all settings, mean squared error (MSE), Bias and Monte Carlo standard error (MCSE) are based on the 1000 simulated datasets.
%\item[$^{2}$] Random censoring\vspace*{6pt}
\end{tablenotes}
\vspace{1cm}  % Add spacing between tables
%%\end{sidewaystable}%[h]
%\end{table*}
%\bigskip\bigskip  % provide some separation between the two tables
%\begin{table*}[t]
%%\begin{sidewaystable} [h]
\caption{Simulation results for scenario 2: $X_i \sim \text{Gamma}(\frac{2}{3},3)$, $W_i \lvert X_i \sim \text{Poisson} (X_i)$, without censoring and 20\% censoring.}
\label{t:tab2}
\tabcolsep=0pt%%
\begin{tabular*}{\textwidth}{@{\extracolsep{\fill}}llcccccc@{\extracolsep{\fill}}}
\hline
& & \multicolumn{3}{@{Mean Estimate}c@{}}{Without Censoring$^{1}$} & \multicolumn{3}{@{Mean Estimate}c@{}}{20\% Random Censoring} \\
\cline{3-5}\cline{6-8}%
Parameter & Methods & Estimator ($\hat{\beta}_x$)  & MSE(MCSE) & Bias (MCSE) &  Estimator ($\hat{\beta}_x$) & MSE(MCSE) & Bias (MCSE) \\
\hline
$\beta_x=0.5$ & Baye-Gamma  & 0.5442    & 0.0118(0.0005) & 0.0442(0.0023)  & 0.532    & 0.0115(0.0007) & 0.0320(0.0032)  \\
& CoxPH.True  & 0.5138    & 0.0049(0.0002) & 0.0138(0.0013)  & 0.5132   & 0.0060(0.0003) & 0.0132(0.0026)  \\
& CoxPH.Naive   & 0.3002    & 0.0435(0.0004) & -0.1998(0.0014) & 0.3004   & 0.0439(0.0006) & -0.1996(0.0020) \\
& DP-Mix-Gamma & 0.5316   & 0.0103(0.0005) & 0.0316(0.0028) & 0.5203 & 0.0105(0.0006)	& 0.0203(0.0037) \\
& POI-Gamma-SIMEX & 0.4342    & 0.0160(0.0004) & -0.0658(0.0029) & 0.4345   & 0.0167(0.0006) & -0.0655(0.0036) \\
\hline
Parameter & Methods & Estimator ($\hat{\beta}_z$)  & MSE(MCSE) & Bias (MCSE) &  Estimator ($\hat{\beta}_z$) & MSE(MCSE) & Bias (MCSE) \\
\hline
$\beta_z=0.1$ & Baye-Gamma  & 0.0978    & 0.0151(0.0007) & -0.0022(0.0020) & 0.1013   & 0.0190(0.0012) & 0.0013(0.0054)  \\
& CoxPH.True   & 0.0973    & 0.0118(0.0005) & -0.0027(0.0020) & 0.1022   & 0.0150(0.0006) & 0.0022(0.0046)  \\
& CoxPH.Naive  & 0.0877    & 0.0129(0.0005) & -0.0123(0.0022) & 0.0909   & 0.0165(0.0009) & -0.0091(0.0051) \\
& DP-Mix-Gamma & 0.0968    & 0.0149(0.0009) & -0.0032(0.0033) & 0.0977 & 0.0198(0.0013) & -0.0023(0.0066) \\
& POI-Gamma-SIMEX & 0.0975    & 0.0179(0.0006) & -0.0025(0.0031) & 0.0991   & 0.0228(0.0012) & -0.0009(0.0062) \\
\hline
\end{tabular*}
\vspace*{0.8cm}
\begin{tablenotes}%
\item Note: Sample size n=100.  
\item[$^{1}$] In all settings, mean squared error (MSE), Bias and Monte Carlo standard error (MCSE) are based on the 1000 simulated datasets.
%\item[$^{2}$] Random censoring\vspace*{6pt}
\end{tablenotes}
%\end{table*}
\end{sidewaystable}%[h]
\clearpage
%%%%Two covariates simulation studies
%\begin{table*}[t]
\begin{sidewaystable} [t]%[h]
\def\~{\hphantom{0}}
\caption{Simulation results for scenario 3: $X_i \sim \text{Gamma}(2,1)$, $W_i \lvert X_i \sim \text{Poisson} (X_i)$, without censoring and 20\% censoring. } 
\label{t:tab3}
\tabcolsep=0pt%%
\begin{tabular*}{\textwidth}{@{\extracolsep{\fill}}llcccccc@{\extracolsep{\fill}}}
\hline
& & \multicolumn{3}{@{Mean Estimate}c@{}}{Without Censoring$^{1}$} & \multicolumn{3}{@{Mean Estimate}c@{}}{20\% Random Censoring} \\
\cline{3-5}\cline{6-8}%
Parameter & Methods & Estimator ($\hat{\beta}_x$)  & MSE(MCSE) & Bias (MCSE) &  Estimator ($\hat{\beta}_x$) & MSE(MCSE) & Bias (MCSE) \\
\hline
$\beta_x=0.5$ & Baye-Gamma  & 0.6362    & 0.0673(0.0047) & 0.1362(0.0102)  & 0.6665   & 0.1029(0.0054) & 0.1665(0.0063)  \\
& CoxPH.True  & 0.5107    & 0.0080(0.0003) & 0.0107(0.0032)  & 0.5153   & 0.0109(0.0004) & 0.0153(0.0026)  \\
& CoxPH.Naive & 0.2111    & 0.0875(0.0016) & -0.2889(0.0028) & 0.2113   & 0.0883(0.0009) & -0.2887(0.0015) \\
& DP-Mix-Gamma & 0.5776&0.0466(0.0033)&0.0776(0.0091) &
0.6108 & 0.0810(0.0059)	& 0.1108(0.0070)\\
& POI-Gamma-SIMEX & 0.3351    & 0.0400(0.0016) & -0.1649(0.0049) & 0.3351   & 0.0418(0.0010) & -0.1649(0.0030) \\
\hline
Parameter & Methods & Estimator ($\hat{\beta}_z$)  & MSE(MCSE) & Bias (MCSE) &  Estimator ($\hat{\beta}_z$) & MSE(MCSE) & Bias (MCSE) \\
\hline
$\beta_z=0.1$ & Baye-Gamma  & 0.099     & 0.0144(0.0004) & -0.0010(0.0037) & 0.1118   & 0.0186(0.0009) & 0.0118(0.0033)  \\
& CoxPH.True   & 0.0972    & 0.0112(0.0004) & -0.0028(0.0033) & 0.1102   & 0.0151(0.0008) & 0.0102(0.0033)  \\
& CoxPH.Naive   & 0.087     & 0.0116(0.0004) & -0.0130(0.0033) & 0.0973   & 0.0147(0.0008) & -0.0027(0.0030) \\
& DP-Mix-Gamma & 0.0981&0.0148(0.0006)&-0.0019(0.0042)&
 0.1119&0.0184(0.0011)&0.0119(0.0036) \\
& POI-Gamma-SIMEX & 0.091     & 0.0142(0.0006) & -0.0090(0.0039) & 0.1032   & 0.0178(0.0011) & 0.0032(0.0033) \\
\hline
\end{tabular*}
\vspace*{0.8cm}
\begin{tablenotes}%
\item Note: Sample size n=100.  
\item[$^{1}$] In all settings, mean squared error (MSE), Bias and Monte Carlo standard error (MCSE) are based on the 1000 simulated datasets.
%\item[$^{2}$] Random censoring\vspace*{6pt}
\end{tablenotes}
%\end{table*}
%\bigskip\bigskip  % provide some separation between the two tables
\vspace{1cm}  % Add spacing between tables
%%\end{sidewaystable}%[h]
%\end{table*}
%\bigskip\bigskip  % provide some separation between the two tables
%\begin{table*}[t]
%%\begin{sidewaystable}
%\begin{table*}[t]
\caption{Simulation results for scenario 4: $X_i \sim \text{Gamma}(10,1)$, $W_i \lvert X_i \sim \text{Poisson} (X_i)$, without censoring and 20\% censoring.}
\label{t:tab4}
\tabcolsep=0pt%%
\begin{tabular*}{\textwidth}{@{\extracolsep{\fill}}llcccccc@{\extracolsep{\fill}}}
\hline
& & \multicolumn{3}{@{Mean Estimate}c@{}}{Without Censoring$^{1}$} & \multicolumn{3}{@{Mean Estimate}c@{}}{20\% Random Censoring} \\
\cline{3-5}\cline{6-8}%
Parameter & Methods & Estimator ($\hat{\beta}_x$)  & MSE(MCSE) & Bias (MCSE) &  Estimator ($\hat{\beta}_x$) & MSE(MCSE) & Bias (MCSE) \\
\hline
$\beta_x=0.5$ & Baye-Gamma  & 0.481     & 0.0040(0.0002) & -0.0190(0.0014) & 0.4886   & 0.0049(0.0002) & -0.0114(0.0025) \\
& CoxPH.True   & 0.5089    & 0.0032(0.0001) & 0.0089(0.0022)  & 0.5096   & 0.0039(0.0002) & 0.0096(0.0019)  \\
& CoxPH.Naive  & 0.1615    & 0.1155(0.0007) & -0.3385(0.0011) & 0.162    & 0.1154(0.0007) & -0.3380(0.0010) \\
& DP-Mix-Gamma & 0.5265 & 0.0078(0.0005) & 0.0265(0.0020) &
 0.5421 & 0.0110(0.0003) & 0.0421(0.0026) \\
& POI-Gamma-SIMEX & 0.2673    & 0.0579(0.0009) & -0.2327(0.0022) & 0.2687   & 0.0576(0.0008) & -0.2313(0.0019) \\
\hline
Parameter & Methods & Estimator ($\hat{\beta}_z$)  & MSE(MCSE) & Bias (MCSE) &  Estimator ($\hat{\beta}_z$) & MSE(MCSE) & Bias (MCSE) \\
\hline
$\beta_z=0.1$ & Baye-Gamma  & 0.0928    & 0.0254(0.0014) & -0.0072(0.0037) & 0.1074   & 0.0317(0.0013) & 0.0074(0.0036)  \\
& CoxPH.True & 0.0984    & 0.0110(0.0004) & -0.0016(0.0028) & 0.1072   & 0.0146(0.0004) & 0.0072(0.0030)  \\
& CoxPH.Naive & 0.0624    & 0.0145(0.0007) & -0.0376(0.0025) & 0.0726   & 0.0168(0.0007) & -0.0274(0.0030) \\
& DP-Mix-Gamma & 0.0935& 0.0257(0.0014) &-0.0065(0.0039)
& 0.1056 & 0.0318(0.0011)	& 0.0056(0.0042)\\
& POI-Gamma-SIMEX & 0.0715    & 0.0223(0.0011) & -0.0285(0.0029) & 0.0845   & 0.0253(0.0011) & -0.0155(0.0042)\\
\hline
\end{tabular*}
\vspace*{0.8cm}
\begin{tablenotes}%
\item Note: Sample size n=100.  
\item[$^{1}$] In all settings, mean squared error (MSE), Bias and Monte Carlo standard error (MCSE) are based on the 1000 simulated datasets.
%\item[$^{2}$] Random censoring\vspace*{6pt}
\end{tablenotes}
%\end{table*}
\end{sidewaystable}%[h]
\clearpage
%%%%lognormal simulation 
\begin{sidewaystable}%[h]
\section*{Web Appendix D}
\def\~{\hphantom{0}}
\caption{Simulation results for misspecification:: $X_i \sim \text{lognormal}(-1.304,1.549)$, $W_i \lvert X_i \sim \text{Poisson} (X_i)$, without censoring and 20\% censoring. } 
\label{t:tab5}
\tabcolsep=0pt%%
\begin{tabular*}{\textwidth}{@{\extracolsep{\fill}}llcccccc@{\extracolsep{\fill}}}
\hline
& & \multicolumn{3}{@{Mean Estimate}c@{}}{Without Censoring$^{1}$} & \multicolumn{3}{@{Mean Estimate}c@{}}{20\% Random Censoring} \\
\cline{3-5}\cline{6-8}%
Parameter & Methods & Estimator ($\hat{\beta}_x$)  & MSE(MCSE) & Bias (MCSE) &  Estimator ($\hat{\beta}_x$) & MSE(MCSE) & Bias (MCSE) \\
\hline
$\beta_x=0.5$ & Baye-Gamma      & 0.6183    & 0.0718(0.0049) & 0.1183(0.0034)  & 0.6064   & 0.0823(0.0104) & 0.1064(0.0059)  \\
& CoxPH.True         & 0.522     & 0.0110(0.0006) & 0.0220(0.0034)  & 0.5234   & 0.0155(0.0011) & 0.0234(0.0022)  \\
& CoxPH.Naive        & 0.3204    & 0.0405(0.0004) & -0.1796(0.0016) & 0.3151   & 0.0439(0.0013) & -0.1849(0.0031) \\
& DP-Mix-Gamma    & 0.563     & 0.0399(0.0034) & 0.0630(0.0041)  & 0.5501   & 0.0421(0.0040) & 0.0501(0.0040)  \\
& POI-Gamma-SIMEX & 0.4737    & 0.0280(0.0014) & -0.0263(0.0032) & 0.4698   & 0.0295(0.0018) & -0.0302(0.0059) \\
\hline
Parameter & Methods & Estimator ($\hat{\beta}_z$)  & MSE(MCSE) & Bias (MCSE) &  Estimator ($\hat{\beta}_z$) & MSE(MCSE) & Bias (MCSE) \\
\hline
$\beta_z=0.1$ &Baye-Gamma      & 0.1082    & 0.0136(0.0006) & 0.0082(0.0044)  & 0.1002   & 0.0158(0.0004) & 0.0002(0.0046)  \\
&CoxPH.True        & 0.1076    & 0.0124(0.0006) & 0.0076(0.0047)  & 0.0977   & 0.0145(0.0005) & -0.0023(0.0042) \\
&CoxPH.Naive         & 0.1009    & 0.0124(0.0006) & 0.0009(0.0040)  & 0.0951   & 0.0144(0.0003) & -0.0049(0.0044) \\
&DP-Mix-Gamma    & 0.107     & 0.0135(0.0008) & 0.0070(0.0044)  & 0.0994   & 0.0153(0.0005) & -0.0006(0.0043) \\
&POI-Gamma-SIMEX & 0.1025    & 0.0146(0.0006) & 0.0025(0.0042)  & 0.0974   & 0.0168(0.0003) & -0.0026(0.0051) \\
\hline
\end{tabular*}
\vspace*{0.8cm}
\begin{tablenotes}%
\item Note: Sample size n=100.  
\item[$^{1}$] In all settings, mean squared error (MSE), Bias and Monte Carlo standard error (MCSE) are based on the 1000 simulated datasets.
%\item[$^{2}$] Random censoring\vspace*{6pt}
\end{tablenotes}
%\end{table*}
%\bigskip\bigskip  % provide some separation between the two tables
%%%%Two covariates simulation studies
\vspace{1cm}  % Add spacing between tables
%%\end{sidewaystable}%[h]
%\end{table*}
%\bigskip\bigskip  % provide some separation between the two tables
%\begin{table*}[t]
%%\begin{sidewaystable}
\caption{Simulation results for misspecification: $X_i \sim \text{lognormal}(0.235,0.957)$, $W_i \lvert X_i \sim \text{Poisson} (X_i)$, without censoring and 20\% censoring.}
\label{t:tab6}
\tabcolsep=0pt%%
\begin{tabular*}{\textwidth}{@{\extracolsep{\fill}}llcccccc@{\extracolsep{\fill}}}
\hline
& & \multicolumn{3}{@{Mean Estimate}c@{}}{Without Censoring$^{1}$} & \multicolumn{3}{@{Mean Estimate}c@{}}{20\% Random Censoring} \\
\cline{3-5}\cline{6-8}%
Parameter & Methods & Estimator ($\hat{\beta}_x$)  & MSE(MCSE) & Bias (MCSE) &  Estimator ($\hat{\beta}_x$) & MSE(MCSE) & Bias (MCSE) \\
\hline
$\beta_x=0.5$ &Baye-Gamma      & 0.5889    & 0.0293(0.0024) & 0.0889(0.0051)  & 0.5802   & 0.0306(0.0025) & 0.0802(0.0067)  \\
& CoxPH.True         & 0.5132    & 0.0061(0.0002) & 0.0132(0.0013)  & 0.5109   & 0.0067(0.0004) & 0.0109(0.0027)  \\
& CoxPH.Naive         & 0.2703    & 0.0566(0.0008) & -0.2297(0.0015) & 0.2714   & 0.0565(0.0009) & -0.2286(0.0018) \\
& DP-Mix-Gamma    & 0.5528    & 0.0208(0.0021) & 0.0528(0.0052)  & 0.5410    & 0.0200(0.0014) & 0.0410(0.0059)  \\
& POI-Gamma-SIMEX & 0.4085    & 0.0208(0.0007) & -0.0915(0.0027) & 0.4082   & 0.0210(0.0008) & -0.0918(0.0028) \\
\hline
Parameter & Methods & Estimator ($\hat{\beta}_z$)  & MSE(MCSE) & Bias (MCSE) &  Estimator ($\hat{\beta}_z$) & MSE(MCSE) & Bias (MCSE) \\
\hline
$\beta_z=0.1$ & Baye-Gamma      & 0.1041    & 0.0161(0.0007) & 0.0041(0.0047)  & 0.102    & 0.0194(0.0013) & 0.0020(0.0038)  \\
CoxPH.True         & 0.0981    & 0.0116(0.0004) & -0.0019(0.0051) & 0.0983   & 0.0145(0.0007) & -0.0017(0.0030) \\
CoxPH.Naive        & 0.0893    & 0.0127(0.0006) & -0.0107(0.0037) & 0.0890    & 0.0159(0.0011) & -0.0110(0.0033) \\
DP-Mix-Gamma    & 0.1027    & 0.0160(0.0008) & 0.0027(0.0048)  & 0.1011   & 0.0185(0.0010) & 0.0011(0.0036)  \\
POI-Gamma-SIMEX & 0.0949    & 0.0168(0.0010) & -0.0051(0.0040) & 0.0953   & 0.0202(0.0015) & -0.0047(0.0039)\\
\hline
\end{tabular*}
\vspace*{0.8cm}
\begin{tablenotes}%
\item Note: Sample size n=100.  
\item[$^{1}$] In all settings, mean squared error (MSE), Bias and Monte Carlo standard error (MCSE) are based on the 1000 simulated datasets.
%\item[$^{2}$] Random censoring\vspace*{6pt}
\end{tablenotes}
%\end{table*}
\end{sidewaystable}%[h]
\clearpage
\begin{sidewaystable}%[h]
\def\~{\hphantom{0}}
\caption{Simulation results for misspecification:: $X_i \sim \text{lognormal}(0.490,0.637)$, $W_i \lvert X_i \sim \text{Poisson} (X_i)$, without censoring and 20\% censoring. } 
\label{t:tab7}
\tabcolsep=0pt%%
\begin{tabular*}{\textwidth}{@{\extracolsep{\fill}}llcccccc@{\extracolsep{\fill}}}
\hline
& & \multicolumn{3}{@{Mean Estimate}c@{}}{Without Censoring$^{1}$} & \multicolumn{3}{@{Mean Estimate}c@{}}{20\% Random Censoring} \\
\cline{3-5}\cline{6-8}%
Parameter & Methods & Estimator ($\hat{\beta}_x$)  & MSE(MCSE) & Bias (MCSE) &  Estimator ($\hat{\beta}_x$) & MSE(MCSE) & Bias (MCSE) \\
\hline
$\beta_x=0.5$ & Baye-Gamma      & 0.6825    & 0.0986(0.0059) & 0.1825(0.0075)  & 0.6922   & 0.1161(0.0052) & 0.1922(0.0079)  \\
& CoxPH.True         & 0.5139    & 0.0086(0.0004) & 0.0139(0.0035)  & 0.5141   & 0.0110(0.0008) & 0.0141(0.0042)  \\
& CoxPH.Naive         & 0.2017    & 0.0935(0.0012) & -0.2983(0.0021) & 0.202    & 0.0939(0.0016) & -0.2980(0.0028) \\
& DP-Mix-Gamma    & 0.6101    & 0.0700(0.0053) & 0.1101(0.0062)  & 0.6214   & 0.0865(0.0043) & 0.1214(0.0073)  \\
& POI-Gamma-SIMEX & 0.3258    & 0.0444(0.0011) & -0.1742(0.0035) & 0.3268   & 0.0454(0.0015) & -0.1732(0.0049) \\
\hline
Parameter & Methods & Estimator ($\hat{\beta}_z$)  & MSE(MCSE) & Bias (MCSE) &  Estimator ($\hat{\beta}_z$) & MSE(MCSE) & Bias (MCSE) \\
\hline
$\beta_z=0.1$ & Baye-Gamma      & 0.0985    & 0.0160(0.0008) & -0.0015(0.0029) & 0.104    & 0.0173(0.0008) & 0.0040(0.0052)  \\
& CoxPH.True         & 0.0987    & 0.0120(0.0005) & -0.0013(0.0025) & 0.1016   & 0.0130(0.0005) & 0.0016(0.0043)  \\
& CoxPH.Naive         & 0.087     & 0.0131(0.0006) & -0.0130(0.0022) & 0.0916   & 0.0135(0.0005) & -0.0084(0.0046) \\
& DP-Mix-Gamma    & 0.0964    & 0.0161(0.0007) & -0.0036(0.0023) & 0.1061   & 0.0171(0.0007) & 0.0061(0.0060)  \\
& POI-Gamma-SIMEX & 0.0909    & 0.0158(0.0007) & -0.0091(0.0025) & 0.0946   & 0.0158(0.0004) & -0.0054(0.0047) \\
\hline
\end{tabular*}
\vspace*{0.8cm}
\begin{tablenotes}%
\item Note: Sample size n=100.  
\item[$^{1}$] In all settings, mean squared error (MSE), Bias and Monte Carlo standard error (MCSE) are based on the 1000 simulated datasets.
%\item[$^{2}$] Random censoring\vspace*{6pt}
\end{tablenotes}
%\end{table*}
%\bigskip\bigskip  % provide some separation between the two tables
%%%%Two covariates simulation studies
%\begin{table*}[t]
\vspace{1cm}  % Add spacing between tables
%%\end{sidewaystable}%[h]
%\end{table*}
%\bigskip\bigskip  % provide some separation between the two tables
%\begin{table*}[t]
%%\begin{sidewaystable}
\caption{Simulation results for misspecification: $X_i \sim \text{lognormal}(2.255,0.309)$, $W_i \lvert X_i \sim \text{Poisson} (X_i)$, without censoring and 20\% censoring.}
\label{t:tab8}
\tabcolsep=0pt%%
\begin{tabular*}{\textwidth}{@{\extracolsep{\fill}}llcccccc@{\extracolsep{\fill}}}
\hline
& & \multicolumn{3}{@{Mean Estimate}c@{}}{Without Censoring$^{1}$} & \multicolumn{3}{@{Mean Estimate}c@{}}{20\% Random Censoring} \\
\cline{3-5}\cline{6-8}%
Parameter & Methods & Estimator ($\hat{\beta}_x$)  & MSE(MCSE) & Bias (MCSE) &  Estimator ($\hat{\beta}_x$) & MSE(MCSE) & Bias (MCSE) \\
\hline
$\beta_x=0.5$ & Baye-Gamma      & 0.4789    & 0.0043(0.0002) & -0.0211(0.0023) & 0.4862   & 0.0057(0.0002) & -0.0138(0.0024) \\
& CoxPH.True         & 0.5101    & 0.0029(0.0002) & 0.0101(0.0019)  & 0.5117   & 0.0043(0.0002) & 0.0117(0.0023)  \\
& CoxPH.Naive         & 0.1567    & 0.1189(0.0006) & -0.3433(0.0009) & 0.1565   & 0.1193(0.0010) & -0.3435(0.0014) \\
& DP-Mix-Gamma    & 0.5284    & 0.0079(0.0004) & 0.0284(0.0031)  & 0.545    & 0.0129(0.0006) & 0.0450(0.0034)  \\
& POI-Gamma-SIMEX & 0.261     & 0.0610(0.0008) & -0.2390(0.0018) & 0.2605   & 0.0621(0.0014) & -0.2395(0.0029) \\
\hline
Parameter & Methods & Estimator ($\hat{\beta}_z$)  & MSE(MCSE) & Bias (MCSE) &  Estimator ($\hat{\beta}_z$) & MSE(MCSE) & Bias (MCSE) \\
\hline
$\beta_z=0.1$ & Baye-Gamma      & 0.0979    & 0.0259(0.0010) & -0.0021(0.0055) & 0.1063   & 0.0330(0.0015) & 0.0063(0.0038)  \\
& CoxPH.True         & 0.0995    & 0.0125(0.0006) & -0.0005(0.0019) & 0.105    & 0.0138(0.0006) & 0.0050(0.0031)  \\
& CoxPH.Naive        & 0.0657    & 0.0149(0.0006) & -0.0343(0.0038) & 0.0725   & 0.0183(0.0010) & -0.0275(0.0033) \\
& DP-Mix-Gamma    & 0.0974    & 0.0263(0.0010) & -0.0026(0.0055) & 0.1072   & 0.0330(0.0015) & 0.0072(0.0041)  \\
& POI-Gamma-SIMEX & 0.0718    & 0.0234(0.0010) & -0.0282(0.0050) & 0.0804   & 0.0281(0.0016) & -0.0196(0.0047)\\
\hline
\end{tabular*}
\vspace*{0.8cm}
\begin{tablenotes}%
\item Note: Sample size n=100.  
\item[$^{1}$] In all settings, mean squared error (MSE), Bias and Monte Carlo standard error (MCSE) are based on the 1000 simulated datasets.
%\item[$^{2}$] Random censoring\vspace*{6pt}
\end{tablenotes}
%\end{table*}
\end{sidewaystable}%[h]
\clearpage
%%%%uniform simulation 
\begin{sidewaystable}%[h]
\def\~{\hphantom{0}}
\caption{Simulation results for misspecification:: $X_i \sim \text{uniform}(0,1.8)$, $W_i \lvert X_i \sim \text{Poisson} (X_i)$, without censoring and 20\% censoring. } 
\label{t:tab9}
\tabcolsep=0pt%%
\begin{tabular*}{\textwidth}{@{\extracolsep{\fill}}llcccccc@{\extracolsep{\fill}}}
\hline
& & \multicolumn{3}{@{Mean Estimate}c@{}}{Without Censoring$^{1}$} & \multicolumn{3}{@{Mean Estimate}c@{}}{20\% Random Censoring} \\
\cline{3-5}\cline{6-8}%
Parameter & Methods & Estimator ($\hat{\beta}_x$)  & MSE(MCSE) & Bias (MCSE) &  Estimator ($\hat{\beta}_x$) & MSE(MCSE) & Bias (MCSE) \\
\hline
$\beta_x=0.5$ & Baye-Gamma      & 0.8507    & 0.5926(0.0364) & 0.3507(0.0248)  & 0.8764   & 0.7588(0.0339) & 0.3764(0.0217)  \\
& CoxPH.True        & 0.5241    & 0.0435(0.0022) & 0.0241(0.0057)  & 0.5101   & 0.0541(0.0018) & 0.0101(0.0069)  \\
& CoxPH.Naive        & 0.1241    & 0.1514(0.0021) & -0.3759(0.0027) & 0.127    & 0.1519(0.0032) & -0.3730(0.0042) \\
& DP-Mix-Gamma    & 0.7124    & 0.4271(0.0223) & 0.2124(0.0181)  & 0.7453   & 0.5733(0.0421) & 0.2453(0.0123)  \\
& POI-Gamma-SIMEX & 0.2099    & 0.1150(0.0033) & -0.2901(0.0047) & 0.2144   & 0.1201(0.0045) & -0.2856(0.0071) \\
\hline
Parameter & Methods & Estimator ($\hat{\beta}_z$)  & MSE(MCSE) & Bias (MCSE) &  Estimator ($\hat{\beta}_z$) & MSE(MCSE) & Bias (MCSE) \\
\hline
$\beta_z=0.1$ & Baye-Gamma      & 0.1135    & 0.0127(0.0005) & 0.0135(0.0037)  & 0.1172   & 0.0170(0.0011) & 0.0172(0.0032)  \\
& CoxPH.True         & 0.1087    & 0.0114(0.0005) & 0.0087(0.0036)  & 0.1124   & 0.0149(0.0012) & 0.0124(0.0029)  \\
& CoxPH.Naive        & 0.1055    & 0.0109(0.0004) & 0.0055(0.0035)  & 0.1079   & 0.0146(0.0011) & 0.0079(0.0027)  \\
& DP-Mix-Gamma    & 0.1119    & 0.0125(0.0004) & 0.0119(0.0035)  & 0.1148   & 0.0171(0.0011) & 0.0148(0.0036)  \\
& POI-Gamma-SIMEX & 0.1053    & 0.0114(0.0004) & 0.0053(0.0036)  & 0.1089   & 0.0156(0.0011) & 0.0089(0.0027) \\
\hline
\end{tabular*}
\vspace*{0.8cm}
\begin{tablenotes}%
\item Note: Sample size n=100.  
\item[$^{1}$] In all settings, mean squared error (MSE), Bias and Monte Carlo standard error (MCSE) are based on the 1000 simulated datasets.
%\item[$^{2}$] Random censoring\vspace*{6pt}
\end{tablenotes}
%\end{table*}
%\bigskip\bigskip  % provide some separation between the two tables
%%%%Two covariates simulation studies
%\begin{table*}[t]
\vspace{1cm}  % Add spacing between tables
%%\end{sidewaystable}%[h]
%%\begin{sidewaystable}
\caption{Simulation results for misspecification: $X_i \sim \text{uniform}(0,4)$, $W_i \lvert X_i \sim \text{Poisson} (X_i)$, without censoring and 20\% censoring.}
\label{t:tab10}
\tabcolsep=0pt%%
\begin{tabular*}{\textwidth}{@{\extracolsep{\fill}}llcccccc@{\extracolsep{\fill}}}
\hline
& & \multicolumn{3}{@{Mean Estimate}c@{}}{Without Censoring$^{1}$} & \multicolumn{3}{@{Mean Estimate}c@{}}{20\% Random Censoring} \\
\cline{3-5}\cline{6-8}%
Parameter & Methods & Estimator ($\hat{\beta}_x$)  & MSE(MCSE) & Bias (MCSE) &  Estimator ($\hat{\beta}_x$) & MSE(MCSE) & Bias (MCSE) \\
\hline
$\beta_x=0.5$ & Baye-Gamma      & 0.6486    & 0.0815(0.0039) & 0.1486(0.0077)  & 0.6638   & 0.0989(0.0039) & 0.1638(0.0063)  \\
& CoxPH.True         & 0.5163    & 0.0107(0.0005) & 0.0163(0.0027)  & 0.513    & 0.0134(0.0004) & 0.0130(0.0038)  \\
& CoxPH.Naive         & 0.1879    & 0.1013(0.0008) & -0.3121(0.0013) & 0.1893   & 0.1012(0.0016) & -0.3107(0.0026) \\
& DP-Mix-Gamma    & 0.5863    & 0.0587(0.0050) & 0.0863(0.0105)  & 0.6037   & 0.0798(0.0056) & 0.1037(0.0084)  \\
& POI-Gamma-SIMEX & 0.3003    & 0.0522(0.0011) & -0.1997(0.0024) & 0.3023   & 0.0533(0.0016) & -0.1977(0.0041) \\
\hline
Parameter & Methods & Estimator ($\hat{\beta}_z$)  & MSE(MCSE) & Bias (MCSE) &  Estimator ($\hat{\beta}_z$) & MSE(MCSE) & Bias (MCSE) \\
\hline
$\beta_z=0.1$ & Baye-Gamma      & 0.1073    & 0.0135(0.0005) & 0.0073(0.0037)  & 0.1089   & 0.0174(0.0009) & 0.0089(0.0024)  \\
& CoxPH.True        & 0.108     & 0.0112(0.0005) & 0.0080(0.0035)  & 0.1121   & 0.0147(0.0011) & 0.0121(0.0028)  \\
& CoxPH.Naive         & 0.0966    & 0.0107(0.0004) & -0.0034(0.0034) & 0.0989   & 0.0145(0.0008) & -0.0011(0.0023) \\
& DP-Mix-Gamma    & 0.1075    & 0.0138(0.0007) & 0.0075(0.0046)  & 0.1129   & 0.0174(0.0009) & 0.0129(0.0032)  \\
& POI-Gamma-SIMEX & 0.0991    & 0.0129(0.0005) & -0.0009(0.0039) & 0.1030    & 0.0173(0.0009) & 0.0030(0.0023) \\
\hline
\end{tabular*}
\vspace*{0.8cm}
\begin{tablenotes}%
\item Note: Sample size n=100.  
\item[$^{1}$] In all settings, mean squared error (MSE), Bias and Monte Carlo standard error (MCSE) are based on the 1000 simulated datasets.
%\item[$^{2}$] Random censoring\vspace*{6pt}
\end{tablenotes}
%\end{table*}
\end{sidewaystable}%[h]

\clearpage
\begin{sidewaystable}%[h]
\def\~{\hphantom{0}}
\caption{Simulation results for misspecification: $X_i \sim \text{uniform}(0,20)$, $W_i \lvert X_i \sim \text{Poisson} (X_i)$, without censoring and 20\% censoring.}
\label{t:tab11}
\tabcolsep=0pt%%
\begin{tabular*}{\textwidth}{@{\extracolsep{\fill}}llcccccc@{\extracolsep{\fill}}}
\hline
& & \multicolumn{3}{@{Mean Estimate}c@{}}{Without Censoring$^{1}$} & \multicolumn{3}{@{Mean Estimate}c@{}}{20\% Random Censoring} \\
\cline{3-5}\cline{6-8}%
Parameter & Methods & Estimator ($\hat{\beta}_x$)  & MSE(MCSE) & Bias (MCSE) &  Estimator ($\hat{\beta}_x$) & MSE(MCSE) & Bias (MCSE) \\
\hline
$\beta_x=0.5$ & Baye-Gamma      & 0.5089    & 0.0016(0.0001) & 0.0089(0.0013)  & 0.506    & 0.0017(0.0001) & 0.0060(0.0010)  \\
& CoxPH.True         & 0.5099    & 0.0025(0.0001) & 0.0099(0.0016)  & 0.5063   & 0.0032(0.0002) & 0.0063(0.0021)  \\
&CoxPH.Naive        & 0.2179    & 0.0804(0.0005) & -0.2821(0.0010) & 0.22     & 0.0795(0.0006) & -0.2800(0.0011) \\
&DP-Mix-Gamma    & 0.507     & 0.0015(0.0001) & 0.0070(0.0011)  & 0.5079   & 0.0018(0.0001) & 0.0079(0.0012)  \\
&POI-Gamma-SIMEX & 0.3254    & 0.0340(0.0007) & -0.1746(0.0021) & 0.3308   & 0.0332(0.0007) & -0.1692(0.0023) \\
\hline
Parameter & Methods & Estimator ($\hat{\beta}_z$)  & MSE(MCSE) & Bias (MCSE) &  Estimator ($\hat{\beta}_z$) & MSE(MCSE) & Bias (MCSE) \\
\hline
$\beta_z=0.1$ & Baye-Gamma      & 0.0996    & 0.0298(0.0015) & -0.0004(0.0055) & 0.1029   & 0.0354(0.0013) & 0.0029(0.0051)  \\
& CoxPH.True        & 0.1049    & 0.0119(0.0006) & 0.0049(0.0040)  & 0.1097   & 0.0154(0.0006) & 0.0097(0.0031)  \\
& CoxPH.Naive         & 0.0626    & 0.0157(0.0008) & -0.0374(0.0040) & 0.065    & 0.0181(0.0006) & -0.0350(0.0029) \\
& DP-Mix-Gamma    & 0.1009    & 0.0295(0.0013) & 0.0009(0.0059)  & 0.1030    & 0.0365(0.0017) & 0.0030(0.0045)  \\
& POI-Gamma-SIMEX & 0.0776    & 0.0314(0.0018) & -0.0224(0.0060) & 0.0813   & 0.0323(0.0011) & -0.0187(0.0042) \\
\hline
\end{tabular*}
\vspace*{0.8cm}
\begin{tablenotes}%
\item Note: Sample size n=100.  
\item[$^{1}$] In all settings, mean squared error (MSE), Bias and Monte Carlo standard error (MCSE) are based on the 1000 simulated datasets.
%\item[$^{2}$] Random censoring\vspace*{6pt}
\end{tablenotes}
\end{sidewaystable}%[h]
\clearpage
%\begin{center}

\begin{table*}%[!h]%
%\begin{sidewaystable}
\section*{Web Appendix E}
\def\~{\hphantom{0}}
%\end{table*}
%\bigskip\bigskip  % provide some separation between the two tables
%\begin{table*}[t]
\caption{Summary statistics of patient characteristics stratified by death event status for COEUR HGSC cohort.}
\label{t:tab12}
 \begin{tabular*}{\textwidth}{@{\extracolsep\fill}llllll@{}}
\hline  
& &Alive & Dead & Total &\\
\hline
&CD3+CD8+FoxP3+ $\sim$E$^{\tnote{*}}$ & & & &\\  
&~ ~ 0  & 159   & 424   & 583 &  \\
&~ ~ $>$ 0   & 50    & 83    & 133 &  \\
&         &       &       &     &  \\
&Age (60+)    & & & &\\
&~ ~ Yes & 119   & 297  & 416  & \\
&~ ~ No  & 90    & 210   & 300  & \\
 &       &       &       &    &   \\
&Non-optimal debulking     & & & & \\
&~ ~ Yes       & 42    & 244   & 286  & \\
&~ ~ No        & 115   & 151   & 266  & \\
&~ ~ NA        & 52    & 112   & 164 & \\
 &         &       &       &     &  \\
&FIGO stage (III/IV)   & & & &\\  
&~ ~ Yes  & 134   & 450   & 584 &  \\
&~ ~ No   & 75    & 57    & 132 &  \\
&         &       &       &     &  \\
\hline
\end{tabular*}
\begin{tablenotes}%%[341pt]
\vspace{0.5cm}
\item[$^{\rm *}$] Indication of the expression of Tregs CD3+CD8+FoxP3+ in the epithelial area
%\item[$^{\rm c}$] Example for a second table footnote.
%\item {\it Source}: Example for table source text.
\end{tablenotes}
\end{table*}
\vspace{0.5cm}
\begin{figure*} [h]
%\section*{Web Appendix E}
\centerline{\includegraphics[scale=0.80]{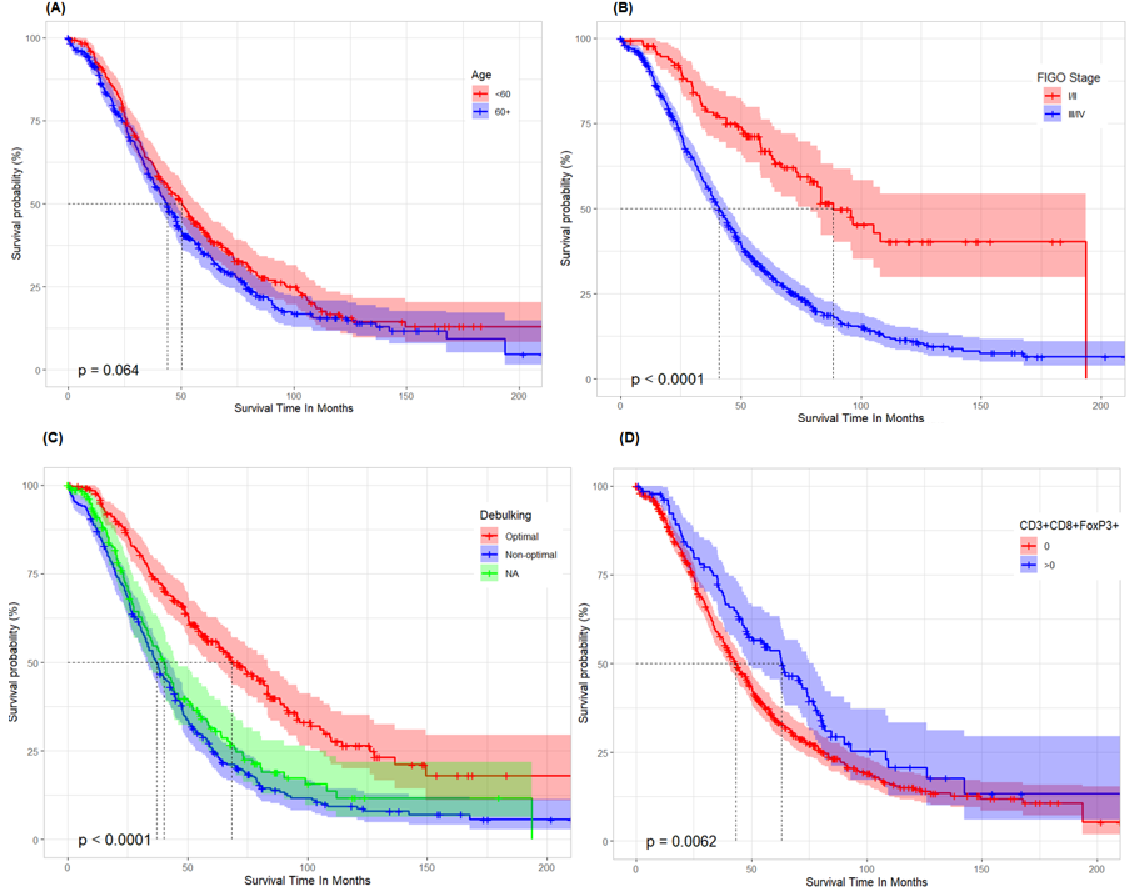}}
\caption{Kaplan-Meier survival curves with pointwise 95\% confidence interval band, log-rank test p-value, and median survival time (dash line) by strata.}
\label{f:sfig6}
\end{figure*}
%\end{sidewaystable}%[h]
%\end{center}
%\begin{center}
%\begin{table*}%[!ht]%
%\begin{sidewaystable}%[h]
\clearpage
\begin{figure*} [h]
\section*{Web Appendix F}
\centerline{\includegraphics[scale=1.1]{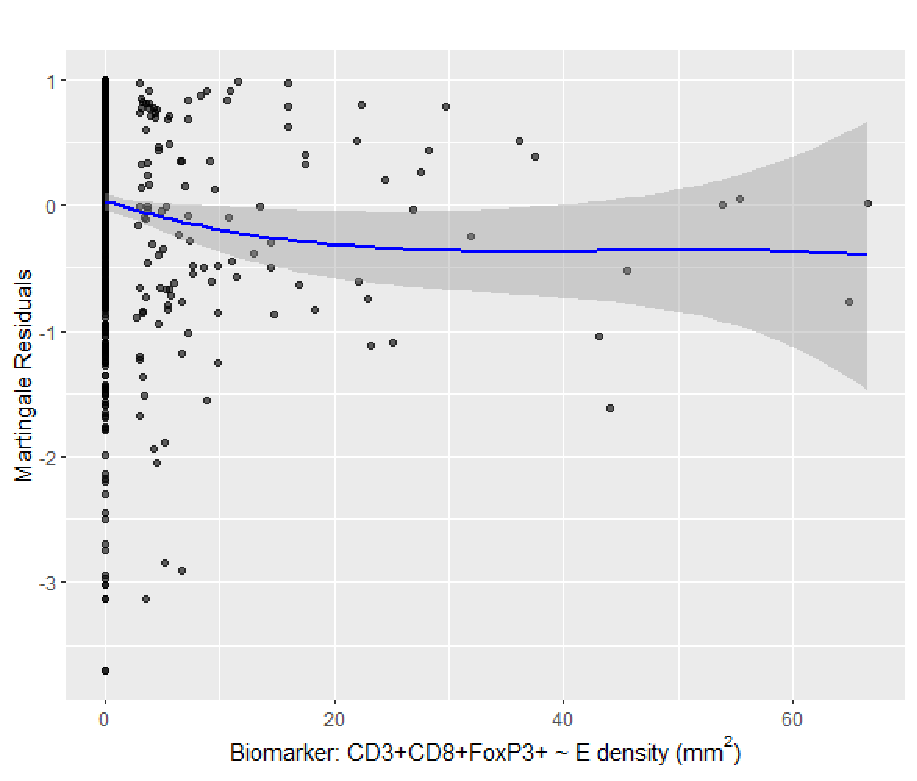}}
\caption{Martingale residuals from the Cox proportional hazards model excluding the biomarker (CD3+CD8+FoxP3+$\sim$ E), plotted against biomarker density. The lack of a nonlinear pattern supports the linearity assumption of the Cox PH model.}
\label{f:sfig7}
\vspace{10em}
\end{figure*}

\begin{table*}%[!h]%
%\begin{sidewaystable}
\section*{Web Appendix G}
\def\~{\hphantom{0}}
\def\~{\hphantom{0}}
\caption{Bayes factor sensitivity analysis for biomarker CD3+CD8+FoxP3+$\sim$E in Dirichlet Process mixture Cox PH model.}
\label{t:tab13}
\begin{tabular*}{\textwidth}{@{\extracolsep\fill}llllllllll@{}}
\hline
Prior for $\beta_x$ & Parameters  & Coef.  & HR  & HR$_{Lower}$ & HR$_{Upper}$ & BF$_{10}^{\tnote{*}}$\\ 
\hline
Normal ($\mu=0$, $\sigma^2=0.01$)  & CD3+CD8+FoxP3+$\sim$E 
&-0.0235& 0.9768 & 0.9505	&1.0017	&0.6624\\
Normal ($\mu=0$, $\sigma^2=1$)  & CD3+CD8+FoxP3+$\sim$E & 
-0.0240	& 0.9763 & 0.9483 & 1.0014& 0.0679 \\
Normal ($\mu=0$, $\sigma^2=10$)& CD3+CD8+FoxP3+$\sim$E& 
-0.0241 & 0.9762 & 0.9489 & 1.0027& 0.0241 \\
Normal ($\mu=0$, $\sigma^2=100$) &CD3+CD8+FoxP3+$\sim$E& -0.0240  & 0.9763 & 0.9491 & 1.0015 &  0.0068  \\
Cauchy ($location=0$, $scale=1$) & CD3+CD8+FoxP3+$\sim$E&
-0.0238	& 0.9765 & 0.9499 & 1.0019 & 0.0549 \\
\hline
\end{tabular*}
\vspace{0.8cm}
\begin{tablenotes}%%[341pt]
\item[$^{\rm *}$] A Bayes factor (BF$_{10}$) of 0.6624 provides anecdotal evidence in favor of the null hypothesis when a highly concentrated null prior is applied.
\end{tablenotes}
\end{table*}
%\end{sidewaystable}%[h]
%\clearpage
%\end{center}
\begin{figure*}
\centerline{\includegraphics[scale=0.80]{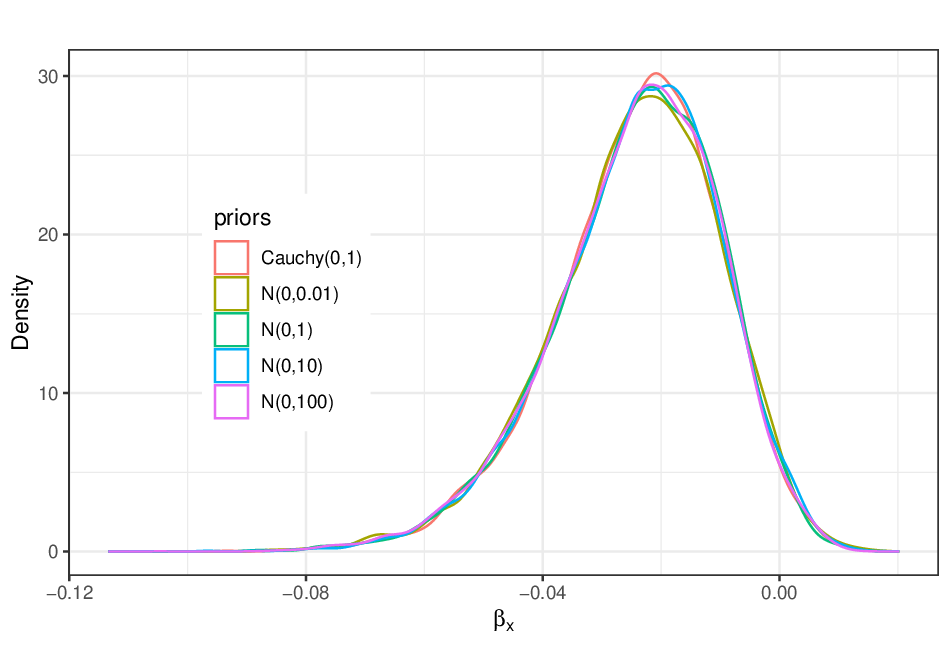}}
\caption{Posterior density estimates for the regression coefficient associated with the CD3+CD8+FoxP3+$\sim$E biomarker $\beta_x$ under the Dirichlet Process mixture Cox proportional hazards model, obtained using varying prior specifications. The resulting densities exhibit negligible sensitivity to the choice of prior, especially when the sample size is sufficiently large.}
\label{f:sfig8}
\end{figure*}